\documentclass[10pt,journal,compsoc]{IEEEtran}

\usepackage{nicefrac}       % compact symbols for 1/2, etc.
\usepackage{epsfig}
\usepackage{graphicx}
\usepackage{amsmath}
\usepackage{amssymb}
\usepackage{url}
\usepackage{amsthm}
\usepackage{algorithm}               %format of the algorithm
\usepackage{algorithmic}             %format of the algorithm
\usepackage{multirow}                %multirow for format of  table
\usepackage{bm}
\usepackage{footmisc}
\usepackage{color}
\usepackage{booktabs}       % professional-quality tables
\usepackage{color}
\usepackage{xfrac}
\usepackage{subfigure}
\usepackage{anyfontsize}
\usepackage[T1]{fontenc}

\ifCLASSOPTIONcompsoc
  \usepackage[nocompress]{cite}
\else
  \usepackage{cite}
\fi
\ifCLASSINFOpdf
\else
\fi
\allowdisplaybreaks

\def\R{\mathbb{R}}

\def\l{\left}
\def\r{\right}
\def\tr{}

\author{%
  David S.~Hippocampus\thanks{Use footnote for providing further information
    about author (webpage, alternative address)---\emph{not} for acknowledging
    funding agencies.} \\
  Department of Computer Science\\
  Cranberry-Lemon University\\
  Pittsburgh, PA 15213 \\
  \texttt{hippo@cs.cranberry-lemon.edu} \\
}
\begin{document}

\title{Rotation Equivariant Proximal Operator for \\Deep Unfolding Methods in Image Restoration}
\author{Jiahong Fu, Qi Xie, Deyu Meng and Zongben Xu
\thanks{This work was supported by the National Natural Science Foundation of China under Grant No. U21A6005, 62206214, 12226004, 62331028 and 62272375.

\noindent
Jiahong Fu, Qi Xie (corresponding author) and Zongben Xu are with School of Mathematics and Statistics and Ministry of Education Key Lab of Intelligent Networks and Network Security, Xi'an Jiaotong University, Shaanxi, P.R.China.

\noindent
Deyu Meng is with School of Mathematics and Statistics and Ministry of  Education Key Lab of  Intelligent Networks and Network Security, Xi’an Jiaotong University,  Xi’an, Shaanxi, China, and Pazhou Laboratory (Huangpu), Guangzhou, Guangdong, China.
}
\thanks{Email:jiahongfu@stu.xjtu.edu.cn, \{xie.qi, dymeng, zbxu\}@mail.xjtu.edu.cn}
%\thanks{Manuscript received April 19, 2019; revised August 26, 2015.}
	%$^\ast${\small Corresponding author}
}

\markboth{IEEE Transactions on Pattern Analysis and Machine Intelligence, 2021}
{Fu \MakeLowercase{\textit{et al.}}: An Equivariant Proximal Operator for Deep Unfolding Methods in Image Restoration}

\IEEEtitleabstractindextext{
\begin{abstract}
The deep unfolding approach has attracted significant attention in computer vision tasks, which well connects conventional image processing modeling manners with more recent deep learning techniques. Specifically, by establishing a direct correspondence between algorithm operators at each implementation step and network modules within each layer, one can rationally construct an almost ``white box'' network architecture with high interpretability. In this architecture, only the predefined component of the proximal operator, known as a proximal network, needs manual configuration, enabling the network to automatically extract intrinsic image priors in a data-driven manner. In current deep unfolding methods, such a proximal network is generally designed as a CNN architecture, whose necessity has been proven by a recent theory. That is, CNN structure substantially delivers the translational symmetry image prior, which is the most universally possessed structural prior across various types of images. However, standard CNN-based proximal networks have essential limitations in capturing the rotation symmetry prior, another universal structural prior underlying general images. This leaves a large room for further performance improvement in deep unfolding approaches. To address this issue, this study makes efforts to suggest a high-accuracy rotation equivariant proximal network that effectively embeds rotation symmetry priors into the deep unfolding framework. Especially, we deduce, for the first time, the theoretical equivariant error for such a designed proximal network with arbitrary layers under arbitrary rotation degrees. This analysis should be the most refined theoretical conclusion for such error evaluation to date and is also indispensable for supporting the rationale behind such networks with intrinsic interpretability requirements. Through experimental validation on different vision tasks, including blind image super-resolution, medical image reconstruction, and image de-raining, the proposed method is validated to be capable of directly replacing the proximal network in current deep unfolding architecture and readily enhancing their state-of-the-art performance. This indicates its potential usability in general vision tasks. The code of our method is available at \url{https://github.com/jiahong-fu/Equivariant-Proximal-Operator}.
\end{abstract}

}
%\mathfrak{}
\maketitle
\IEEEdisplaynontitleabstractindextext
\IEEEpeerreviewmaketitle

\newtheorem{Thm}{Theorem}
\newtheorem{Rem}{Remark}
\newtheorem{Lem}{Lemma}
\newtheorem{Cor}{Corollary}

\IEEEraisesectionheading{\section{Introduction}\label{sec:introduction}}

\IEEEPARstart{W}{ith} the rapid development of deep learning (DL),
convolutional neural network-based approaches have achieved great successes in the field of image restoration (IR). Among them, the deep unfolding approach finely integrates the advantages of conventional model-driven and current data-driven DL methodologies. This has recently attracted extensive research attention and achieved state-of-the-art (SOTA) performance over multiple IR tasks\cite{barbu2009training}, \cite{samuel2009learning},\cite{sun2011learning}, \cite{yang2016deep, zhang2020deep}, e.g., blind image super-resolution \cite{fu2022kxnet}, medical image reconstruction \cite{wang2021dicdnet}, and image de-raining \cite{wang2020model}.

The superiority of the deep unfolding approach mainly lies in its two specific characteristics. On one hand, it explicitly associates network modules in each layer with operators at each step of a carefully designed algorithm. This results in a deep unfolding network with an almost ``white box'' architecture, where only the priori solution part of the proximal operator \cite{romano2017little, cohen2021regularization} is required to be manually pre-specified, resulting in a high degree of interpretability. This facilitates the network to easily embed helpful domain knowledge of a specific IR task, especially its underlying physical generalization mechanism, into its architecture construction. This then naturally makes the network better customized and more properly designed against the investigated task, and achieves good performance in diverse image processing fields, like natural image restoration and medical imaging reconstruction \cite{barbu2009training, samuel2009learning, sun2011learning, yang2016deep, zhang2020deep}.

On the other hand, the priori solution component of the proximal operator, as the main learning component of a deep unfolding network, facilitates the capability of the network to extract intrinsic image priors in a data-driven manner automatically. Specifically, it has been widely recognized that the performance of traditional model-driven methods heavily depends on  meticulously pre-designed prior terms, with widely used priors such as sparsity, low-rankness and smoothness. However, manually crafting image priors is an invariably challenging endeavor, especially for real images with complex configurations. In contrast, the deep unfolding approach explores a way of learning image priors through a parameterized proximal operator (i.e., a proximal network) exclusively from data. This significantly enhances the model's ability to represent image priors and thereby boosts its overall performance for general IR tasks.

As the most essential component, a proper design for such a proximal network always plays a critical role in the final effect of the deep unfolding methods. Currently, a series of proximal network forms have been attempted on different IR tasks, such as ResNet-based \cite{he2016deep, xie2020mhf}, U-Net-based \cite{ronneberger2015u, zhang2017learning, zhang2020plug}, and transformer-based \cite{zhang2022practical} proximal networks. Actually, an interesting observation is that most studies lean to select certain CNN-based network structures to design such proximal network forms. The reasoning behind this inclination and its necessity are illuminated by a recent theory expounded in \cite{celledoni2021equivariant}. That is, the CNN architecture can be proved to possess substantial a translation equivariance\footnote{Translation equivariance means that shifting an input image of CNN is equivalent to shifting all of its intermediate feature maps and output image, and translation invariance means that shifting an input image of CNN does not change its intermediate feature maps and output image.} property in theory, intrinsically delivering the translational symmetry image prior. This symmetry prior is universally present in general images, like natural, medical and remote sensing image domains. Succeeded from conventional prior terms, most of which are necessary with translational invariance property, CNN naturally becomes the most rational selection for setting proximal network in deep unfolding methodology.

\begin{figure}
    % \vspace{1mm}
    \centering
    \includegraphics[width=\linewidth, scale=1.0]{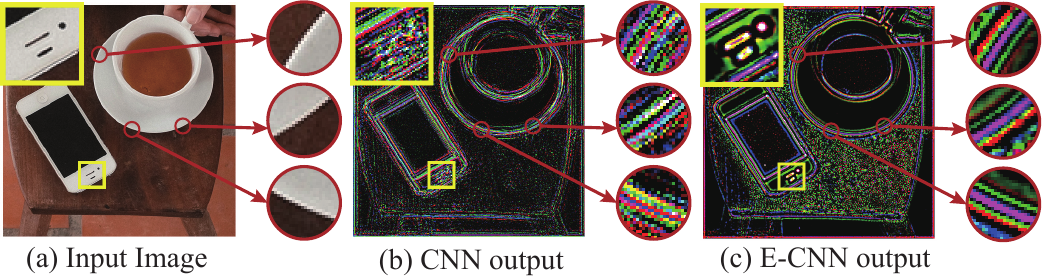}
    \caption{Illustration of the output feature map of a typical image obtained by standard CNN and our used rotation equivariant convolution neural network (E-CNN).}
    \label{fig:fig1}
    \vspace{-2mm}
\end{figure}

\begin{figure}[t]
    % \vspace{1mm}
    \centering
    \includegraphics[width=\linewidth, scale=1.0]{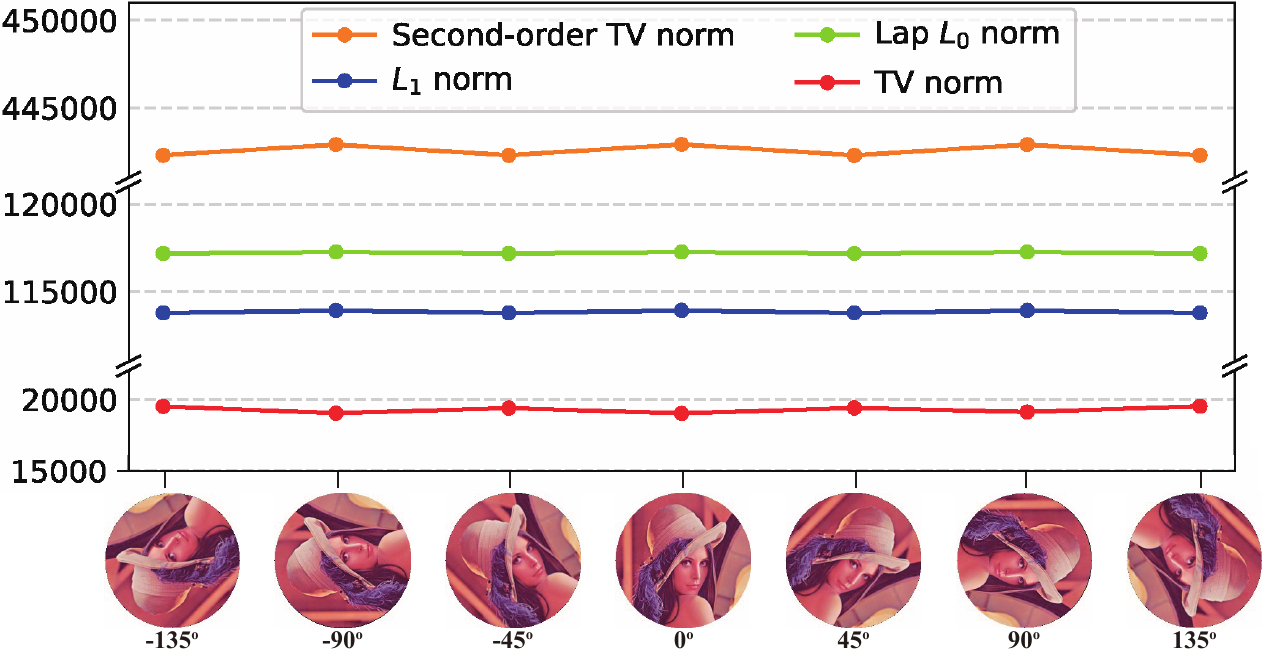}
    \caption{The values of four typical conventional regularization terms on the Lena image with different rotation angles.}
    \label{fig:figlena7}
    \vspace{-2mm}
\end{figure}

The main value of the theory presented in \cite{celledoni2021equivariant} is that it provides a fundamental principle for purposefully designing proximal networks in deep unfolding implementations, i.e., embedding more meaningful and general transformation symmetry priors underlying the investigated IR task into the network. From this perspective, the currently used proximal networks still have evident limitations and performance improvement room. The most typical one is that current standard CNN-based proximal networks are limited in capturing the rotation symmetry prior, which yet should be another universal structural prior underlying general images besides translation symmetry. As intuitively demonstrated in Fig. \ref{fig:fig1}(a), different image patches in an image can exhibit similar patterns, though they are in different orientations. Actually, rotation symmetry prior can be rational embedded in the rotation invariance\footnote{Rotation invariance means that rotating the input image of an operator will not change the output.}. It is easy to deduce that most of the traditionally used regularization terms for reflecting image priors are invariant with respect to orientation, as well as translation. For example, rotating an image wouldn't theoretically change the values of the commonly used local smoothness, non-local self-similarity, low-rankness or other image prior terms. Fig. \ref{fig:figlena7} demonstrates this regularizer property in general, where one can easily observe that the values of Lap $L_0$ \cite{shen2006image}, $L_1$ \cite{elad2010sparse}, TV and second-order TV norms \cite{rudin1992nonlinear} of the same image only vary slightly when the image is rotated\footnote{Their mild intensity variations are conducted by discretization error of image representation. More dataset statistics can be found in the supplementary material.}. According to the principle provided in \cite{celledoni2021equivariant}, this implies that orientation equivariance of the proximal network for IR tasks should also be necessarily desired, which yet not hold for current mainstream CNNs, as clearly shown in Fig. \ref{fig:fig1}(b).

In this study, we devote to exploiting rotation equivariant convolution network and theory, aiming to make it capable of being properly equipped on proximal network design for general IR tasks. The method is expected to be both easily operated in practice and strictly supported in theory, where the latter is especially important and indispensable for the deep unfolding methodology that has inherent interpretability requirement.

Specifically, the main contribution of this paper can be mainly summarized as follows:

1) We suggest a high-accuracy rotation equivariant proximal network for deep unfolding methods through comprehensive attempts and evaluations. By easily replacing the convolutions in current proximal networks with such rotation equivariant amelioration, the rotational symmetry image prior can be readily embedded into the deep unfolding network, and also help make the network more lightweight by allowing convolutions able to share representations for more essentially similar local structures.

2) We provide thorough theoretical equivariant error analysis for the suggested rotation equivariant proximal network with arbitrary layers under arbitrary rotation degrees, which strictly reveals how the selected rotation group’s cardinality for equivariant convolutions and the image resolution influence the equivariant error. To the best of our knowledge, this should be the most refined theoretical conclusion for such error evaluation to date, and thus meaningful for general deep unfolding network design.

3) By readily replacing the proximal components within current deep unfolding networks with the suggested rotation equivariant ones for different IR tasks, including blind image super-resolution, medical image reconstruction and image deraining, we have substantiated that such amelioration facilitates the model able to consistently surpass the performance of SOTA methods designed for these tasks and also exhibit superior generalization capabilities. This indicates the potential applicability of this method across more general deep unfolding techniques.

The remainder of this paper is organized as follows. Section 2 provides an overview of related works and introduces the concept of equivariant prior knowledge and definitions. Section 3 presents the suggested equivariant proximal operator network with its error analysis theory. Section 4 demonstrates experiments on evaluating the performance of the method on general IR tasks. The paper finally concludes with a future
work discussion.

\section{Related Work}
% \subsection{Deep Neural Network for Image Restoration}
% In recent years, learning-based image restoration methods \cite{dong2014learning, xu2014deep, zhang2017learning}, particularly those based on CNNs, have become incrasingly popular due to their superior performance compared to traditional model-based methods \cite{buades2005non, dabov2007image, yang2010image}.

\subsection{Deep Unfolding Network for Image Restoration}
Deep unfolding methods can be mainly categorized into deep Plug-and-Play (PnP) and deep unrolling methods, which have been widely used for various IR tasks.

The early PnP algorithms can be traced back to \cite{zoran2011learning, danielyan2011bm3d, heide2014flexisp, venkatakrishnan2013plug, rond2016poisson}. By replacing the proximal operator in Alternating Direction Method of Multipliers (ADMM) with off-the-shelf denoisers (e.g., BM3D \cite{dabov2007image} or Total Variation \cite{rudin1992nonlinear}), Venkatakrishnan et al. \cite{venkatakrishnan2013plug} proposed the concept of PnP priors. These innovative approaches inspire researchers to replace the explicit prior of model-based methods with an implicit prior of denoiser for PnP methods. Such advantages lay the foundation for leveraging deep CNN denoisers to improve effectiveness. Recently, deep PnP methods \cite{zhang2017learning, meinhardt2017learning, rick2017one, dong2018denoising} have achieved good progress across multiple tasks by introducing a learnable denoiser with deep learning. These works \cite{zhang2017learning, tirer2018image, zhang2017beyond, he2018optimizing} have applications such as denoising, deblurring, super-resolution, inpainting, and medical imaging. Compared with PnP, which requires a manually specified denoiser, deep PnP inclines to learn a more accurate data-related denoiser prior through a deep neural network, which not only shortens the steps of the algorithm but also achieves better performance.

To leverage domain knowledge and observation model for image restoration, deep unrolling methods aim to replace certain steps in an algorithm with learnable parameters \cite{sun2011learning, gregor2010learning, afonso2010fast}. Inspired by LISTA \cite{gregor2010learning}, Wang et al. \cite{wang2015deep} designed a deep unrolling method for image super-resolution based on spare coding. In \cite{yang2016deep}, Yang et al. proposed ADMM-Net, which is derived from the iterative procedures in alternating direction method of multipliers algorithm for optimizing a compressive-sensing-based magnetic resonance imaging (MRI) model. Since then, a series of deep unfolding methods based on specific optimization algorithms have been applied to various IR tasks.
Such algorithm unfolding attempts include first-order-gradient-based unfolding networks \cite{chen2016trainable, adler2017solving, putzky2017recurrent,sprechmann2013supervised, yang2016deep, rick2017one}, second-order-gradient-based unfolding network \cite{xiong2013supervised, lilearning} and Half-Quadratic-Splitting (HQS) algorithm based unfolding method \cite{yang2018proximal, zhang2017learning, zhang2020deep} and so on. Typically, there is a class of Iterative Shrinkage-Thresholding Algorithms (ISTA)  \cite{blumensath2008iterative, beck2009fast}, which is usually with simple iterative formulations for IR tasks, and has been shown to be suitable for designing deep unfolding networks \cite{moreau2017understanding, xie2020mhf, wang2020model, aberdam2021ada}.

In the aforementioned algorithms (e.g. AMDD, HQS, and ISTA), the proximal operator usually plays an important role. Intrinsically, such proximal operator is related to the pre-specified prior terms, e.g., the $L_1$-norm term, representing the sparsity prior, corresponds to a soft-thresholding operator. Thus, proximal network design is the key issue in constructing unfolding networks. The most intuitive way \cite{gregor2010learning, liu2019alista} for proximal operator design is to set scalars/matrices \textit{etc.} in iterative algorithms as learnable parameters and learn them by data-driven training. These methods usually need to manually assume a prior for the correlated tasks, and then deduce the formulation of the corresponding thresholding operator before training. To release the difficulty and inaccuracy in manually designing image prior, more latest researchers attempt to further parameterize the proximal operator as a pure  deep neural network, i.e., proximal network, based on Convolution+BN+ReLu \cite{aljadaany2019proximal}, Encoder-Decoder structure \cite{rick2017one, zhang2020deep}, ResNet structure \cite{adler2018learned, xie2020mhf, wang2020model} and so on. Recently, transformer-based proximal networks have also received attention \cite{zhang2022practical}.

The key drawback of these data-dependent proximal network attempts is that most of them are heuristically constructed with existing deep network architectures, while rarely evaluating their underlying construction mechanism on how they could insightfully embed intrinsic image priors. This research focuses on the most typical issue along this research line, i.e., how to embed rotation symmetry prior, one of the most universal priors underlying general images, into a proximal network to enhance the capability of deep unfolding in an implementable easy and theoretically sound manner.

\subsection{Group Equivariant CNNs}
A series of recent works have been proposed to incorporate equivariance into networks, for modeling
the transform symmetry priors existing in data samples, such as translation, rotation, scaling and reflection symmetries.
Formally, let $\Psi$ be a mapping from the input feature space to the output feature space, and $G$ be a group of transformations. We say $\Psi$ is group equivariance about $G$, if for any $g \in G$,
\begin{equation}
    \Psi \left [ \pi_{g} [X]\right] = \pi_g' \left[ \Psi [X]\right]
    \label{eq:2.1}
\end{equation}
where $X$ represents an input image or a feature map, and $\pi_{g}$ and $\pi_g'$ denote how the transformation $g$ acts on input and output features, respectively.

%Early exploration of transform symmetry priors for images was mainly through heuristic ways. One of the most common and intuitive methods is data augmentation \cite{krizhevsky2017imagenet}, which aims to improve the robustness of the model to transformations by adding transformed training data. In addition, there is a transformation-invariant pooling operator \cite{laptev2016ti} or differentiable modules \cite{esteves2017polar} to achieve equivariant transformation.

Multiple pieces of literature have begun to investigate how to explicitly incorporate transformation equivariance into neural network architectures. Specifically, G-CNN \cite{cohen2016group} and  HexaConv \cite{hoogeboom2018hexaconv} explicitly integrate $\sfrac{\pi}{2}$ and  $\sfrac{\pi}{3}$ degree rotation  equivariances into the neural network, respectively. Subsequently, a more comprehensive equivariance is achieved based on interpolation \cite{zhou2017oriented, marcos2017rotation} and Gaussian-resampling \cite{worrall2017harmonic} technology. Very recently, the filter parametrization technique has been employed for designing G-CNNs. Early attempts \cite{weiler2018learning, weiler2019general} were made to use harmonics as steerable filters to achieve exact equivariance with respect to larger transformation groups in the continuous domain. After that, \cite{shen2020pdo} and \cite{shen2021pdo} designed equivariance by relating convolution with partial differential operators and proposed PDO-eConv.
However, these methods still suffer from low expression accuracy problems in filter parametrization methods, which results in their inferior performance in IR tasks.
Very recently, Xie et al. \cite{xie2022fourier}  proposed Fourier
series expansion-based filter parametrization, which has relatively high expression accuracy. They further constructed F-Conv, an equivariant convolution method suitable for IR tasks, and it has been verified to be effective.

The main issue of current rotation equivariant convolution research should be on their theoretical level, which can only provide equivariant error support on limited discrete orientation degrees on one single network layer. This issue is particularly critical for deep unfolding context since this methodology places a specific emphasis on network interpretability. Therefore, it particularly expects a clear theoretical error evaluation to support the rationale of embedding rotation equivariance into multi-layer proximal networks repetitively appearing throughout the deep unfolding network. This is also the main focus of this study, i.e., ensuring that the implementation of rotation equivariant embedding is not only heuristic but also provides valuable scientific inspiration for future research.

% \begin{figure}
%     \centering
%     \includegraphics[width=\linewidth, scale=1.0]{./images/lena_rot_5.pdf}
%     \caption{Displayed the rotation invariance of image priors.}
%     \label{fig:figlena5}
% \end{figure}

\section{Equivariant Proximal Network}\label{sec:sec3}

% For convolutional networks, the feature maps and convolutional kernels can be naturally modeled as functions defined in the continuous domain. As \cite{xie2022fourier} constructed parameterized convolutions for input layer $\Psi$, intermediate layer $\Phi$, and output layer $\Upsilon$. Specifically, input layer $\Psi$ denotes the convolution imposed on the input feature maps, which maps an input $r \in C^{\infty}(\mathbb{R}^2)$ to a feature map defined on $E(2)$. Intermediate layer $\Phi$ denotes the convolution on an intermediate layer, which maps a feature map $e \in C^{\infty} (E(2))$ to another feature map defined on $E(2)$. Output layer $\Upsilon$ denotes the convolution on the final layer, which maps a feature map $e\in C^{\infty}(E(2))$ to a function in $C^{\infty}(\mathbb{R}^2)$.

% \begin{Thm}\label{Thm1}
%     For $r \in C^{\infty} (\mathbb{R}^2)$, $y \in \mathbb{R}^2$, $\tilde{A} \in O(2)$, the following results are satisfied:
%     \begin{equation}
%         \Upsilon \left[\Phi \left[\Psi \left[\pi_{\tilde{A}}^{R} \left[r\right]\right]\right]\right] (y) = \pi_{\tilde{A}}^{R} \left[\Upsilon \left[\Phi \left[\Psi \left[r\right]\right]\right]\right] (y)
%     \end{equation}
%     where $\pi_{\tilde{A}}^{R}(y) = r(\tilde{A}^{-1} y), \forall y \in \mathbb{R}^2$.
% \end{Thm}
\subsection{General Framework of Deep Unfolding }
For a general image restoration task, the underlying high-quality image ${X}$ is expected to be restored from its degraded measurement ${Y}$. Formally, the correlated ill-posed inverse problem can be formulated as following \cite{engl1996regularization}:
\begin{equation}
  \min_{{X}} L({X},{Y}) + \lambda {R}({X}),
  \label{eq:eq1}
\end{equation}
where $\lambda$ denotes the trade-off parameter;  $L({X},{Y})$ is the data fidelity term, which is often set as $L({X},{Y})=\left \| {Y} - {A} {X}\right \|_F^2$; and ${R}({X})$ is the regularization term.
It should be noted that in traditional model-based IR methods, the design of  ${R}({X})$ plays the most substantial role in image prior modeling and significantly influences application performance. Correspondingly, designing ${R}({X})$  is also one of the most challenging tasks in IR method formulations.

There are multiple ways for solving Eq. (\ref{eq:eq1}). In this paper, we mainly consider the ISTA algorithm \cite{blumensath2008iterative, beck2009fast}, which should be one of the most popular choices for addressing general IR tasks. Generally, in the ISTA algorithm, the following quadratic approximation of (\ref{eq:eq1}) is iteratively solved:
\begin{equation}
\begin{split}
     \min_{{X}}  L\left({X}^{(t)},{Y}\right)&+\left\langle\nabla L\left({X}^{(t)}, {Y}\right), {X}-{X}^{(t)} \right\rangle\\
     &+\frac{1}{2\eta}\left\|{X}-{X}^{(t)}\right\|_F^2+\lambda R({X}),
\end{split}
\label{eq:ista}
\end{equation}
where ${X}^{(t)}$ denotes the updating result after the $t^{th}$ iteration. After organizing and simplifying the updating equation, we can obtain the following update process:
\begin{equation}
    {X}^{(t+1)} \!=\! \arg \min_{{X}} \frac{1}{2}\left\| {X} - \tilde {{X}}^{(t)}   \right\|_F^2 \!+\! \eta\lambda \mathcal{R}({X}),
    \label{eq:ista_simplify}
\end{equation}
where $ \tilde {{X}}^{(t)} = {X}^{(t)} - \eta \nabla h\left({X}^{(t)}\right) $ and $\eta$ denotes the step-size parameter.

Solving the regularized minimization problem (\ref{eq:ista_simplify}) is the key progress of the algorithm. Usually, the following
 close-form solutions can be deduced\cite{donoho1995noising}:
\begin{equation}
        {X}^{(t+1)} \!= \! \operatorname{prox}_{R}\left(\tilde {{X}}^{(t)}\right) \!= \! \operatorname{prox}_{R} \left( {X}^{(t)}  \!- \! \eta \nabla h \l({X}^{(t)}\r)\right),
    \label{eq:ista_prox}
\end{equation}
where $\operatorname{prox}_{R}(\cdot)$ is the proximal operator \cite{moreau1962fonctions} corresponding to $R(\cdot)$.
It is easy to find that the image prior knowledge is indeed mainly reflected in $\operatorname{prox}_{R}(\cdot)$ when solving the problem. This inspires the construction of deep unfolding networks, aiming to directly encode image prior in its proximal network modules. This approach eliminates the need for the manual design of regularization terms and solving of the regularized minimization problem (\ref{eq:ista_simplify}), simultaneously.
In the past years, several deep-leaning-based proximal networks have been constructed to characterize such data-dependent priors \cite{yang2016deep} \cite{zhang2017learning} \cite{xie2020mhf}, which well employs such a data-driven prior learning manner and achieved good performance across various IR tasks.

%----------------------------------------------
\begin{figure*}[t]
   \centering
   \includegraphics[width=1\linewidth]{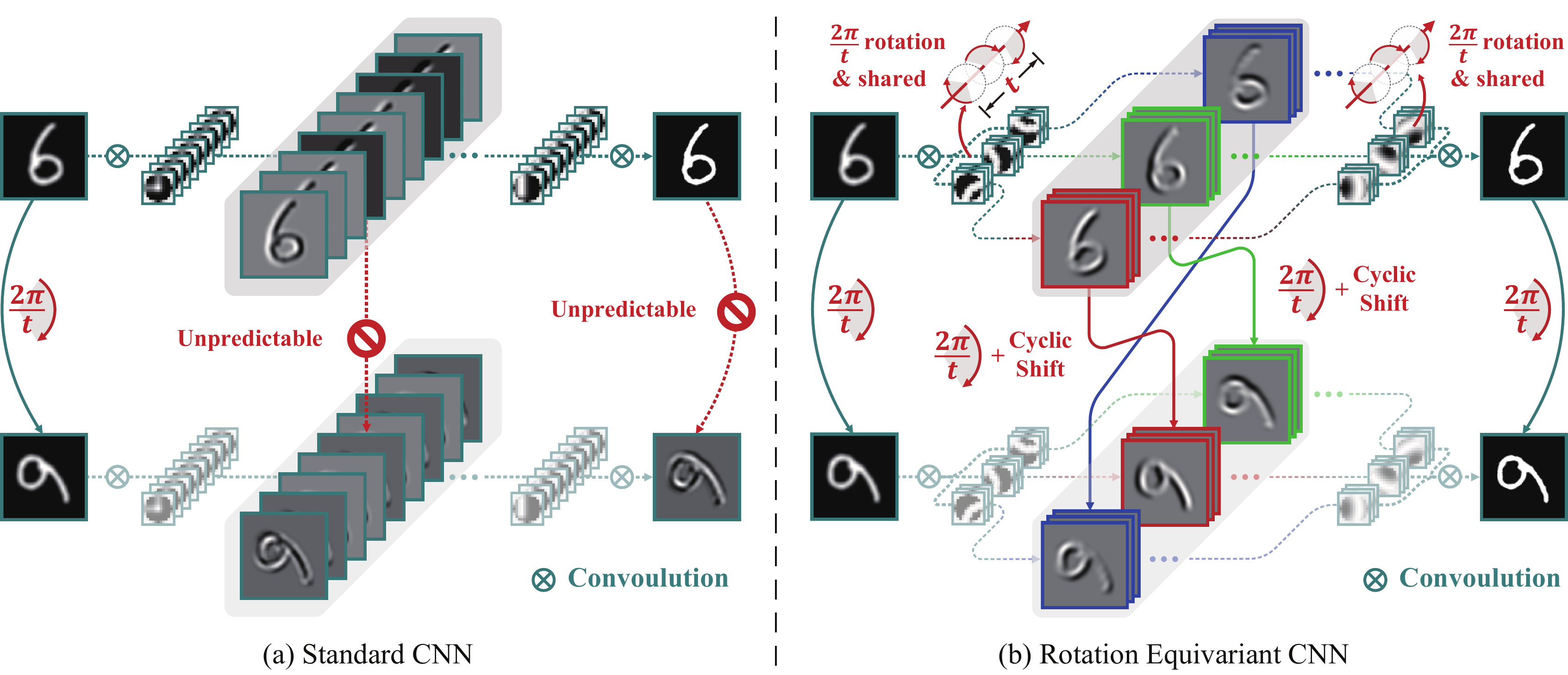}
   \caption{Illustrations of the feature maps and outputs of different CNNs when rotating the input images for $\nicefrac{2k\pi}{t}$, where $k = 1,2,\cdots, t$, $t$ is the equivariant number. (a) A standard CNN. (b) The rotation equivariant CNN with the same backbone. }
   \label{fig:equva_proximal}
   \vspace{-1mm}
\end{figure*}
%----------------------------------------------

\subsection{A Principle for Proximal Network Design}

As aforementioned, the performance of a deep unfolding method highly relies on the rational design of the proximal network, which plays an implicit role in incorporating crucial image prior knowledge into the network and thus facilitating easy training and good generalization of the entire network. Most of the existing deep unfolding methods tend to use multi-layer CNNs/ResNet to specify this proximal network \cite{yang2016deep,zhang2017learning,xie2020mhf}, but less considering why this architecture is necessary and how to further incorporate more important image priors into it.

Very Recently, Celledoni et al. \cite{celledoni2021equivariant} conducted a meaningful exploration against this issue, and revealed a principle for such proximal network $\operatorname{prox}_R(\cdot)$ design. Specifically, it is theoretically proved that when ${R}(\cdot)$ is invariant to certain transformations like translation or rotation, then the proximal operator must be equivariant to the transformation. \tr{ Formally, 
regarding an image as an element in a Hilbert space, the following lemma can be deduced \cite{celledoni2021equivariant}.}
\tr{
\begin{Lem}
Suppose that $\mathcal{X}$ is a Hilbert space and $\pi_g: \mathcal{X} \rightarrow \mathcal{X}$ is a unitary transformation of a group $G$ on $\mathcal{X}$. 
If a functional $R: \mathcal{X} \rightarrow \mathbb{R}\cup \{+\infty\}$  has a well-defined single-valued proximal operator $\operatorname{prox}_R: \mathcal{X} \rightarrow \mathcal{X}^*$, and $R$ is invariant on $\mathcal{X}^*$, i.e. $R(\pi_g({X})) = R({X})$, $\forall {X}\in \mathcal{X}^*$, 
then $\operatorname{prox}_{R}$ is equivariant, in the sense that
    \begin{equation}
        \operatorname{prox}_R[\pi_g]({X}) = \pi_g[\operatorname{prox}_R]({X}), \forall {X}\in\mathcal{X}, g\in G.
    \label{eq:eq3}
    \end{equation}
\end{Lem}
}
According to Lemma 1, it can be deduced that a key principle in designing a rational proximal network is to ensure its equivariance to certain unitary transformations when the regularization term for the target image is invariant under these transformations. \tr{Lemma 1 also indicates that  the equivariance of the proximal operator is only related to the regularization invariance of the output image of the proximal operator (i.e., the underlying high-quality image). Therefore, even if the input image undergoes degradation that disrupts its rotational symmetry,  rendering it non-invariant under regularization, the necessity for rotation equivariance in the proximal network  remains rational.}

This conclusion actually explains why convolution-based proximal networks (such as CNNs and ResNets) are effective. Specifically, since images usually exhibit evident \emph{translation symmetry}, it is easy to find that most conventional regularization terms designed for image data are with strict translation invariant property, i.e.,  $R[\pi_t]({X}) = R({X})$, \tr{where $\pi_t$ denotes the translation transformation.}
. Moreover, $\pi_t$ is indeed a unitary transformation\footnote{Translation of a digital image ${X}$ can be performed by changing the positions of elements of ${X}$, while changing element positions is a unitary transformation.}. This satisfies the condition of Lemma 1. Therefore, according to Celledoni's theory, translation equivariance is necessary for ISTA algorithms and their corresponding deep unfolding networks. As the most typical translation equivariant operator \cite{kondor2018generalization}, the convolution structure naturally becomes a rational choice for setting proximal networks.

\subsection{Rotation Equivariant Proximal Network}

Besides translation symmetry, rotation symmetry should also be considered as a universal structural prior underlying general images. As shown in Fig. \ref{fig:figlena7},
%In Fig. \ref{fig:dataset_rot_invariant},
%the rotation invariance of multiple regularization terms, $R(\cdot)$, are evaluated via
%the histogram of the relative rotation invariant error $\frac{R\l[\pi_\theta\r](X) - \operatorname{mean}_\theta\l[R\l[\pi_\theta\r]\r](X)}{\operatorname{mean}_\theta\l[R\l[\pi_\theta\r]\r](X)}$, $\theta \sim U(-\pi,\pi)$ on a real image dataset Flickr2K\footnote{more dataset statistics can be found in the supplementary material., timofte2017ntire}.
it is evident that the relative rotation invariant errors of typical commonly used regularization terms are consistently small. This implies that conventional regularization terms always have essentially embedded such important rotation invariance knowledge inside, also validating the universality of this specific image invariance prior. Consequently, this inspires us to necessarily design a rotation equivariant proximal network for deep unfolding networks based on Lemma 1.

It should be noted that most of the current proximal networks can indeed not be proved to be rotation equivariant. As shown in Fig. \ref{fig:equva_proximal}(a), in a standard convolutional network, when the input image is rotated, the relationship between feature maps before and after rotating the input image is intrinsically unpredictable.
Fortunately, multiple rotation equivariant convolutions have been proposed in the past years, including GCNN \cite{cohen2016group}, E2-CNN \cite{weiler2019general}, PDO-eConv \cite{shen2020pdo} and F-Conv \cite{xie2022fourier}. These rotation equivariant convolutions demonstrate the potential for constructing a rotation equivariant proximal network for IR tasks.

Specifically, by replacing all convolutions in the CNN-based proximal networks with rotation equivariant ones, we can obtain rotation equivariant proximal networks without changing other architectures of the backbone\footnote{\tr{If there is batch normalization operator in the network, it should also be replaced with rotation equivariant ones \cite{weiler2018learning}.}}, as shown in Fig. \ref{fig:equva_proximal}.
For a network based on equivariant convolutions, when the input image is rotated by $\nicefrac{2k\pi}{t}$ degrees ($k = 1,2,\cdots, t$, where $t$ is the equivariant number of the equivariant convolutions, as intuitively depicted in Fig. \ref{fig:equva_proximal} (b) for easy understanding), then the change of all the feature maps are predictable. This change results from the combination of rotation and cyclic shifts in channels, and the output of the network is also rotated for $\nicefrac{2k\pi}{t}$ degree. One can view Fig. \ref{fig:equva_proximal} to easily understand this progress.

Besides, by embedding such rotation equivariance to the CNN proximal network, the network is possibly expected to be more lightweight than commonly used CNNs. From Fig. \ref{fig:equva_proximal} (a) and (b), one can easily see that the convolution filters in rotation equivariant convolutions have been rotated for $\frac{2k\pi}{t}$,  $k = 1,2,\cdots,t$ degrees before performing convolution operator, which means that all parameters for parameterized filters are reused $t$ times. Therefore, when using the same number of channels, the rotation equivariant convolution only requires $\nicefrac{1}{t}$ convolution filters and thus tends to have fewer network parameters. This inclination to make the network parameters more efficiently used also hopefully leads to better generalization capabilities.

%------------------------------------
\begin{table*}[t]
    % \footnotesize
    \renewcommand\arraystretch{1.0}
      \centering
            \caption{Average PSNR/SSIM of different competing methods on synthesized testing datasets.} %\vspace{-2mm}
               \setlength{\tabcolsep}{15pt}
      \begin{tabular}{c c c c c c c c c c }
         \toprule
        \multirow{2}{*}{Method}  & \multicolumn{2}{c}{Urban100 \cite{huang2015single}} & \multicolumn{2}{c}{BSD100 \cite{martin2001database}} & \multicolumn{2}{c}{Set14 \cite{zeyde2010single}} & \multicolumn{2}{c}{Set5 \cite{bevilacqua2012low}}   \\
        \cmidrule(r){2-3}\cmidrule(r){4-5} \cmidrule(r){6-7}\cmidrule(r){8-9}
         & PSNR & SSIM & PSNR & SSIM & PSNR & SSIM & PSNR & SSIM  \\
         \midrule
        $\operatorname{Prox_{CNN}}$  & 30.57 & 0.8877 & 29.56 & 0.8171 & 29.86 & 0.8112 & 31.84 & 0.8807 \\
        $\operatorname{Prox_{SCUNet}}$ & 30.50 & 0.8887 & 29.55 & 0.8176 & 29.96 & \bf 0.8159 & 31.79 & 0.8805\\
        $\operatorname{Prox_{G-CNN}}$  & 30.75 & 0.8914 & 29.60  & 0.8199 & 29.89 & 0.8111 & 31.91 & 0.8825 \\
        $\operatorname{Prox_{E2-CNN}}$  & 30.66 & 0.8891 & 29.58 & 0.8177 & 29.79 & 0.8086 & 31.90 & 0.8817 \\
        $\operatorname{Prox_{PDO-eConv}}$ & 29.55 & 0.8628 & 29.19 & 0.7998 & 29.50 & 0.8030 & 31.47 & 0.8703 \\
        $\operatorname{Prox_{F-Conv}}$  & \bf 31.05 & \bf 0.8962 & \bf 29.66 & \bf 0.8211 & \bf 30.02 & 0.8134 & \bf 31.98 & \bf 0.8836 \\
       \bottomrule

      \end{tabular}
    % \vspace{0.1cm}
      \label{tab:table01}
\end{table*}
%------------------------------------------------------------------------

\begin{figure*}[ht]
\vspace{-2mm}
\centering
\subfigure{
    \begin{minipage}[b]{0.12\linewidth}
    \centerline{\small Noisy Image} \vspace{0.7mm}
    \centerline{\includegraphics[width=2.4cm]{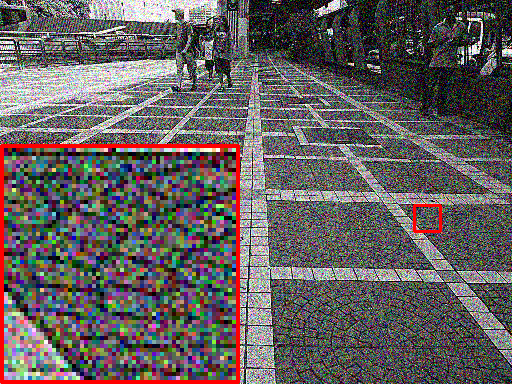}}
    \centerline{\small PSNR/SSIM}
    \end{minipage}
    } %[图片大小]{图片路径}
\subfigure{
    \begin{minipage}[b]{0.12\linewidth}
    \centerline{\small $\operatorname{Prox_{CNN}}$}\vspace{0.7mm}
    \centerline{\includegraphics[width=2.4cm]{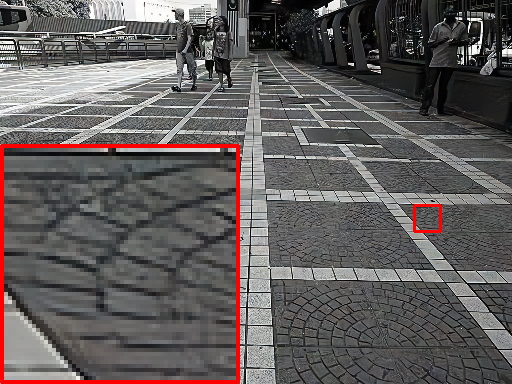}}
    \centerline{\small 25.38/0.7916}
    \end{minipage}
    } %[图片大小]{图片路径}
\subfigure{
    \begin{minipage}[b]{0.12\linewidth}
    \centerline{\small $\operatorname{Prox_{SCUNet}}$}\vspace{0.7mm}
    \centerline{\includegraphics[width=2.4cm]{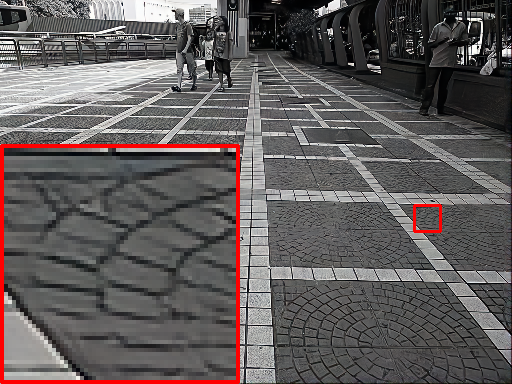}}
    \centerline{\small 25.29/0.7836}
    \end{minipage}
    } %[图片大小]{图片路径}
\subfigure{
    \begin{minipage}[b]{0.12\linewidth}
    \centerline{\small $\operatorname{Prox_{G-CNN}}$}\vspace{0.7mm}
    \centerline{\includegraphics[width=2.4cm]{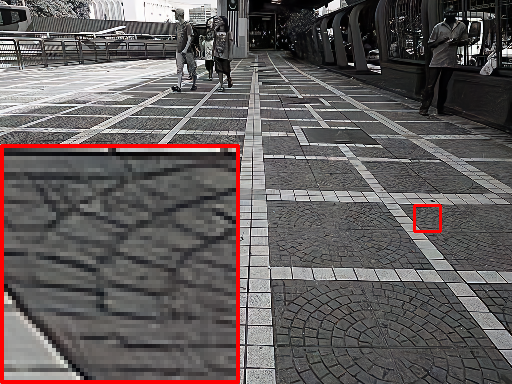}}
    \centerline{ \small 25.38/0.7934}
    \end{minipage}
    } %[图片大小]{图片路径}
\subfigure{
    \begin{minipage}[b]{0.12\linewidth}
    \centerline{\small $\operatorname{Prox_{E2-CNN}}$}\vspace{0.7mm}
    \centerline{\includegraphics[width=2.4cm]{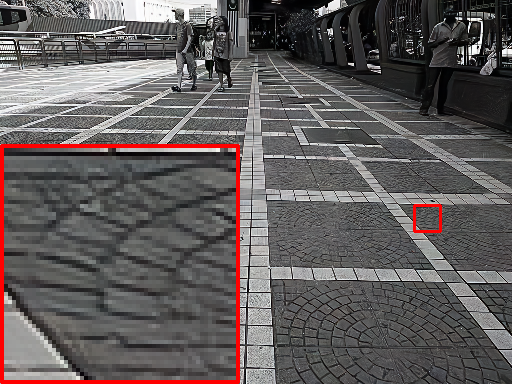}}
    \centerline{\small 25.37/0.7930}
    \end{minipage}
    } %[图片大小]{图片路径}
\subfigure{
    \begin{minipage}[b]{0.12\linewidth}
    \centerline{\small $\operatorname{Prox_{PDO-e}}$}\vspace{0.7mm}
    \centerline{\includegraphics[width=2.4cm]{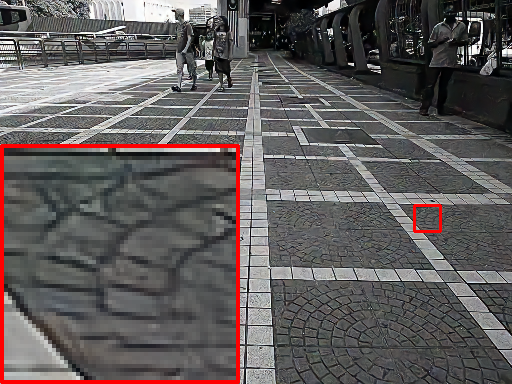}}
    \centerline{\small 24.80/0.7565}
    \end{minipage}
    } %[图片大小]{图片路径}
%\subfigure{
%    \begin{minipage}[b]{0.12\linewidth}
%    \centerline{\small $\operatorname{Prox_{F-Conv}}$}\vspace{0.7mm}
%    \centerline{\includegraphics[width=2.4cm]{images/denoising/result_fconv3x3.png}}
%    \centerline{\small\textbf{25.51}/\textbf{0.7978}}
%    \end{minipage}
%    } %[图片大小]{图片路径}
\subfigure{
    \begin{minipage}[b]{0.12\linewidth}
    \centerline{\small $\operatorname{Prox_{F-Conv}}$}\vspace{0.7mm}
    \centerline{\includegraphics[width=2.4cm]{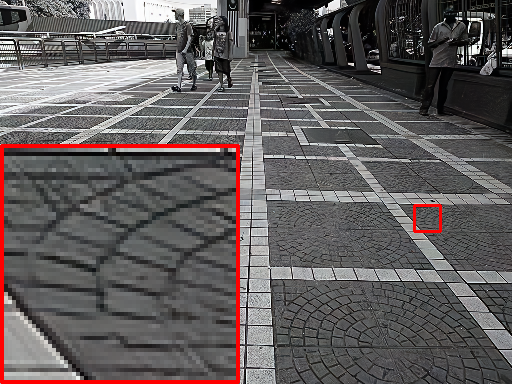}}
    \centerline{\small\textbf{25.58}/\textbf{0.8021}}
    \end{minipage}     } %[图片大小]{图片路径}
% \subfigure[Jackson Yee]{\includegraphics[width=3.5cm]{image/014GT.png}}
% \subfigure[Jackson Yee]{\includegraphics[width=3.5cm]{image/014GT.png}}
\caption{Performance comparison \textit{img 095} in Urban100 \cite{huang2015single}, where the  Gaussian noise is with standard deviation of 50.} %图片标题
\vspace{-2mm}
\label{fig:denoise_vis095}  %图片交叉引用时的标签
\end{figure*}
%------------------------------------

Now, we need to discuss two critical issues to guarantee a sound construction and utilization of such an expected rotation equivariant proximal network in general deep unfolding frameworks. The first is to suggest a proper equivariant convolution representation for implementing the intended task, and the second more important issue is to provide theoretical support to the suggested implementation manner.
Specifically, the current analysis about equivariant error is only on one single convolution layer and with respect to finite discrete rotation degrees $\theta = \nicefrac{2k\pi}{t}$ ($k = 1,2,\cdots, t$). However, the interpretation requirement of deep unfolding methodology calls for the theoretical representation error support in proximal networks at arbitrary angles and multi-layer convolution networks.

\subsection{Selection of Equivariant Convolutions}

In previous pieces of literature \cite{ronneberger2015u, zhang2017learning, xie2020mhf}, the proximal network is always regarded as an image denoiser for the solver of Eq. (\ref{eq:ista_simplify}). It can be similarly seen as a denoising model.
To evaluate the necessity of rotation equivariance for the proximal network and select a proper equivariant convolution for general IR tasks, we have carefully designed simulation denoising experiments for this purpose.

Specifically, the competing methods include standard CNN, Swin-Conv UNet (SCUNet) \cite{zhang2022practical} and four SOTA equivariant convolution ameliorations, G-CNN \cite{cohen2016group}, E2-CNN \cite{weiler2018learning}, PDO-eConv \cite{shen2020pdo} and F-Conv \cite{xie2022fourier}. We exploit the same backbone network for competing convolution methods, i.e., a ResNet \cite{he2016deep} containing 16 residual blocks of 256 channels.
We simulate Gaussian noise and train the proximal networks for denoising. All the competing methods use a filter size of $5\times 5$. We set the equivariant number to $t = 4$ for G-CNN since it is designed with $t = 4$ and set $t = 8$ for the other equivariant convolutions.
All methods are trained with 800 samples from the DIV2K dataset \cite{agustsson2017ntire}, and the simulated Gaussian noise has a standard deviation of 50.
Four well-known benchmark image datasets, including Set5 \cite{bevilacqua2012low}, Set14 \cite{zeyde2010single}, BSD100 \cite{martin2001database}, and Urban100 \cite{huang2015single}, are selected as testing datasets for our experiments.

%All the method is trained with 800 samples from DIV2K dataset \cite{agustsson2017ntire}  and test on Urban100 (100 testing samples), BSD100 \cite{martin2001database}, Set14 \cite{zeyde2010single}, and Set5 \cite{bevilacqua2012low} respectively.
%Besides, we test the performance of all the competing convolution method under two filter size settings, i.e., $3\times 3$ and $5\times 5$, excepted for PDO-eConv, for it is only designed for filters with size of $5\times 5$.

As demonstrated in Table \ref{tab:table01}, the F-Conv based method exhibits clear superiority in performance when compared to the standard convolution ($\operatorname{Prox_{CNN}}$) and Swin transformer based network ($\operatorname{Prox_{SCUNet}}$). This superiority can be attributed to the explicit representation of the rotation symmetry prior and the representation accuracy of the filter parametrization in F-Conv based convolutions.
For the E2-CNN and PDO-eConv methods, although the rotation symmetry prior has also been represented,
their performances do not exhibit significant advantages over standard CNN-based methods. This should be due to the limitation of representation accuracy of their filter parameterizations. Moreover, we provide visual denoising results in Fig. \ref{fig:denoise_vis095}, and it is seen that F-Conv based proximal network inclines to restore the rotation symmetry structure better than others.
These results inspire us to suggest F-Conv as the main tool to construct our rotation equivariant proximal network for deep unfolding against general IR tasks.

\tr{
It should be noted that although equivariant convolutions are more complex than regular convolutions, the flops and inference time of F-Conv are actually almost the same as regular convolutions, which implies the practicality of rotation equivariant networks. For more details on the computational complexity of F-Conv, please refer to the supplementary material.}

\subsection{Theoretical Analysis for Equivariant CNN}

%Current equivariant convolutions aim to achieve rotation equivariant with respect to $t$ degrees uniform selected on $[0,2\pi]$, as shown in Fig. \ref{fig:equva_proximal}(b)   \cite{cohen2016group, weiler2019general, shen2020pdo, xie2022fourier}.
It has been well proved that current equivariant convolutions are exact in the continuous domain. However, they would become approximate after necessary discretization for real-world applications \cite{cohen2016group, weiler2019general, shen2020pdo, xie2022fourier}.
In the case of discretization, the latest theoretical analysis has derived the equivariant error specifically for $t$ selected orientations within a single convolution layer, stated as follows\cite{xie2022fourier}:

\begin{Lem} For a one-layer equivariant convolution operation $\operatorname{Conv}_{eq}(\cdot)$  with $t$ being its equivariant number, and an input image (or a feature map) $X$, under proper conditions, the following result holds:
\begin{equation}\label{privious}
\begin{split}
& \left|\operatorname{Conv}_{eq} [\pi_{\theta}] \left(X\right) - {\pi}_{\theta}[\operatorname{Conv}_{eq}]\left( X \right)
    \right|\leq  {C}(p+1)^2h^2, \\
    &\forall \theta = {2k\pi}/{t},~k = 1,2,\cdots,t,
\end{split}
\end{equation}
where  ${\pi}_{\theta}$ defines the rotation transformation of the input image  (or on a feature map\footnote{ When ${\pi}_{\theta}$ is the transformation on a feature map, it is combined of $\theta$ degree rotation and $k$ cyclic shift on the channels of the feature map. Please see more details in the supplementary material. }), $h$ is the mesh size, $p$ is the filter size and $C$  is a positive constant independent of $h,p$.
\end{Lem}

The left of (\ref{privious}) represents the equivariant error, i.e., the difference between the results obtained with rotation before and after convolutions.
Lemma 2 shows that the equivariant error for a single-layer convolution would approach zero as the mesh size $h$ approaches zero when the rotation degree is within the selected discrete set.
This is consistent with human intuition because the equivariant error is caused by discretization, and an error caused by discretization approaches zero as the mesh size approaches zero.

However, it is important to understand how to evaluate the equivariant error for the entire equivariant network under arbitrary rotation degrees. When the rotation angle does not belong to the discrete rotation set or the equivariant convolution involves multiple layers, Lemma 2 can not be used to analyze the equivariant error correspondingly. Specifically,
%as illustrated in Fig. \ref{fig:unaligned},
if the input image is rotated by $\theta \neq \nicefrac{2k\pi}{t}$, then the change of feature maps at each layer becomes unpredictable.
This unpredictability arises because the cyclic shift of channels can only be performed for finite rotation angles $\theta = \nicefrac{2k\pi}{t}$, $k = 1,2,\cdots, t$. Therefore, the current theoretical analysis framework is no longer applicable for such purposes.

In order to deal with this issue, we deduce the following theorem for more comprehensively and accurately evaluating the rotation equivariant error.

%----------------------------------------------
%\begin{figure}[t]
%   \centering
%   \includegraphics[width=1\linewidth]{images/Fig4_network_small.pdf}
%   \caption{When the rotation degree does not belong to the discrete rotation set. The feature maps of the group equivariant convolution can not be aligned.}
%   \label{fig:unaligned}
%\end{figure}
%----------------------------------------------

%\begin{Thm}\label{Thm1}
%For a $n$-layer rotation equivariant CNN network $\operatorname{CNN}_{eq}(\cdot)$ with $t$ to be its equivariant number, and an input image  $X$, under  proper conditions,  the following results holds  for arbitrary rotation degree $\theta$,
%    \begin{equation}
%    \begin{split}
%      \l|\operatorname{CNN}_{eq}\left[\pi_{\theta} \right]\! ({X}) \!-\! \pi_{\theta} \left[\operatorname{CNN}_{eq} \right]\!({X})\r|
%    \!\leq\! C\l(h^2\!\!+\! { p h}{t^{-\!1}}\r)\!,
%    \end{split}
%    \end{equation}
%    where  ${\pi}_{\theta}$ defines the rotation transformation of input image  $p$ is the filter size, $h$ is the mesh size, and $C$ is  a finite constant.
%\end{Thm}

\begin{Thm}\label{Thm1}
For an image ${X}$ with size $H\times W\times n_0$, and a $N$-layer rotation equivariant CNN network $\operatorname{CNN}_{eq}(\cdot)$, whose channel number of the $i^{th}$ layer is $n_i$, rotation equivariant subgroup is $S\leqslant O(2)$, $|S|=t$, and activation function is set as ReLU. If the latent continuous function of the $c^{th}$ channel of ${X}$ denoted as $r_c: \mathbb{R}^2 \! \rightarrow \! \mathbb{R}$, and the latent continuous function of any convolution filters in the $i^{th}$ layer denoted as $\varphi^{i}: \mathbb{R}^2 \! \rightarrow \! \mathbb{R}$, where $i \in \{1, \cdots\!, N\}$, $c \in \{1, \cdots\!, n_{0}\}$, for any $x\in \R^2$, the following conditions are satisfied:
    \begin{equation}
        \begin{split}
            & |r_c(x)| \leq F_0, \|\nabla r_c(x)\| \leq G_0, \|\nabla ^2 r_c(x)\| \leq H_0, \\
            & |\varphi^{i}(x)| \leq F_i, \|\nabla \varphi^{i}(x)\| \leq G_i, \|\nabla ^2 \varphi^{i}(x)\| \leq H_i,\\
            &\forall \|x\|\geq\nicefrac{(p+1)h}{2},~ \varphi_{i}(x)=0,\\
        \end{split}
    \end{equation}
    where $p$ is the filter size, $h$ is the mesh size, and $\nabla$ and ${\nabla}^2$ denote the operators of gradient and Hessian matrix, respectively. For an arbitrary $0 \leq \theta \leq 2 \pi$, $A_{\theta}$ denotes the rotation matrix, the following result is satisfied:
\begin{equation}\label{main_conclusion}
\begin{split}
  \l|\operatorname{CNN}_{eq}\left[\pi_{\theta} \right]\! ({X}) \!-\! \pi_{\theta} \left[\operatorname{CNN}_{eq} \right]\!({X})\r|
\!\leq\! C_1 h^2\!+\! C_2 { p h}{t^{-1}},
\end{split}
\end{equation}
where
 \begin{equation}
    \begin{split}
        & C_1 = 2 N \mathcal{F} \cdot \sum_{i=1}^{N} \! \! \left( \! \frac{H_i F_0}{F_i} \! + \! 2 \frac{G_i}{F_i} \! \sum_{m=1}^{i-1} \! \frac{G_m F_0}{F_m} \! + \! 2 \frac{G_i G_0}{F_i} + H_0 \! \right), \\
        & C_2 = 2 \pi G_0 \mathcal{F} \l(2 \operatorname{max}\{H,\!W\} p^{-1} \! + \! 2 N\r), \mathcal{F} \!= \!\prod \limits_{i=1}^N n_{i-1} p^2 F_i. \\
    \end{split}
\end{equation}
\end{Thm}

It should be indicated that to the best of our knowledge, Theorem \ref{Thm1} should firstly provide the bound of the equivariant error in relation to the multi-layer network under arbitrary rotation degrees\footnote{One can refer to the supplementary material for the details of the proof.}. It reveals that the equivariant error is primarily influenced by two factors: the mesh size ($h$) and the equivariant number ($t$). Consistent with Lemma 2, the equivariant error in Theorem 1 will become larger when the mesh size $h$ increases.
Besides, we can further achieve more insightful conclusions from this theoretical conclusion as follows:

(I) When the range of a convolution kernel is a constant, then $ph$ remains constant. In this scenario, it becomes evident that the second term of the equivariant error bound primarily depends on $t$, and a larger $t$ results in a smaller equivariant error. This fully complies with the intuition that designing equivariant convolution to be equivariant to more selected degrees reduces the equivariant error under arbitrary degrees.

(II) Interestingly, when $t=1$, the network degenerates to a conventional CNN, and the equivariant convolution will not be equivariant anymore. This occurs because the second term in the error bound becomes a constant, and the error bound will never approach zero. This reveals why conventional CNNs do not possess intrinsic rotation equivariance properly.

Table \ref{tab:equi_error} shows the mean rotation equivariant error ($\nicefrac{\l\|f\l[\pi_\theta\r](X) - \pi_\theta[f](X)\r\|_2}{\|f(X)\|_2}$, $\theta \sim U(-\pi,\pi)$) on a typical benchmark image dataset DIV2K\cite{agustsson2017ntire}. We can clearly see that as the equivariant number increases, the equivariant error of the rotation equivariant convolution decreases rapidly. When $t = 1$ the equivariant error is similar to that of the ordinary CNN. The measure of equivariant error here is the relative calculation error. These results verify our aforementioned theoretical conjectures.

(III) Moreover, it is interesting to note that a component, $n_{i-1}p^2$ (contained in the fundamental term of $\mathcal{F}$), consistently emerges in both $C_1$ and $C_2$. This observation fully accords with some well-known initialization schemes for convolution networks, such as Xavier initialization \cite{pmlr-v9-glorot10a} and He's initialization \cite{He_2015_ICCV}. In these initialization schemes, the elements of convolution kernels, denoted as $v^i_\varphi$ at the $i^{th}$ layer, are recommended to be initialized by sampling from Gaussian distribution $N(0,\frac{c}{n_{i-1}p^2})$ or Uniform distribution $U(-\frac{c}{n_{i-1}p^2}, \frac{c}{n_{i-1}p^2})$, where $c$ is a constant number. In other words,  $n_{i-1}p^2v^i_\varphi\sim N(0,c) $ or $U(-c,c)$, which is rational to be assumed to be bounded with high probability. From this perspective, it should be rational to consider $n_{i-1} p^2 F_i$ and $\mathcal{F}$ as bounded combined with the smoothness of the kernel.

%----------------------------
\begin{table}[t]
    \centering
    \caption{The rotation equivariant error, $\nicefrac{\l\|f\l[\pi_\theta\r](X) - \pi_\theta[f](X)\r\|_2}{\|f(X)\|_2}$, averaged over 10 random selected $\theta$ and 100 random selected images on the dataset DIV2K\cite{agustsson2017ntire}.}
    \setlength{\tabcolsep}{7.5pt}\vspace{-1.5mm}
    \begin{tabular}{c c c c c c c }
    \toprule
    \multirow{2}{*}{CNN}  & \multicolumn{6}{c}{Rotation Equivariant CNN} \\    \cmidrule(r){2-7}
    & $t=1$ & $t=2$ & $t=4$ & $t=8$ & $t=12$ & $t=24$ \\
    \midrule
        0.851 & 0.852 & 0.648 & 0.428 & 0.259 & 0.175  & 0.068\\
    \bottomrule
    \end{tabular}
    \label{tab:equi_error}\vspace{-4mm}
\end{table}
%----------------------------

In summary, it can be found that all conditions of Theorem \ref{Thm1} are easy to be satisfied in practice, which only need the first and second derivatives of the underlying function of the input image and convolution kernels to be bounded. As a result, This implies that the rotation equivariance of the suggested proximal network can be theoretically guaranteed for arbitrary rotation degrees and multiple network layers, under controllable rotation equivariant error in practice. Consequently, this enables deep unfolding networks equipped with such rotational equivariance proximal network to have solid theoretical guarantees and to maintain their substantial requirements for interoperability.

%----------------------------------------
\begin{table*}[ht]
%\vspace{-6mm}
%\scriptsize
% \small
% \renewcommand\arraystretch{0.90}
  \centering % \setlength{\tabcolsep}{40pt}
    \caption{Average PSNR/SSIM of the super-resolution results obtained by all comparison methods on different benchmark datasets.}
    \centering \setlength{\tabcolsep}{11pt}
\begin{tabular}{l c c c c c c c c c c}
    \toprule
    % \hline
     \multirow{2}{*}{Method} & \multirow{2}{*}{Scale} & \multicolumn{2}{c}{Urban100 \cite{huang2015single}} & \multicolumn{2}{c}{BSD100 \cite{martin2001database}} & \multicolumn{2}{c}{Set14 \cite{zeyde2010single}} & \multicolumn{2}{c}{Set5 \cite{bevilacqua2012low}}& Standard  \\
     \cmidrule(r){3-4}\cmidrule(r){5-6}\cmidrule(r){7-8}\cmidrule(r){9-10}
     & & PSNR & SSIM & PSNR & SSIM & PSNR & SSIM & PSNR & SSIM& Deviation \\
    % \midrule
    \midrule
    % \hline
    % \hline
     Bicubic & \multirow{9}{*}{x2} & 23.00 & 0.6656 & 25.85 & 0.6769 & 25.74 & 0.7085 & 27.68 & 0.8047 &\multirow{27}{*}{0} \\
%     RCAN \cite{zhang2018image} & & 23.22 & 0.6791 & 26.03  & 0.6896 & 25.92 & 0.7217 & 27.85 & 0.8095 &\\
     IKC \cite{gu2019blind} & & 27.46 & 0.8401 & 29.85 & 0.8390 & 30.69 & 0.8614 & 33.99 & 0.9229 &\\
     DASR \cite{wang2021unsupervised} & & 26.65 & 0.8106 & 28.84 & 0.7965 & 29.44 & 0.8224 & 32.50 & 0.8961 & \\
     DAN \cite{luo2020unfolding} & & 27.93 & 0.8497 & 30.09 & 0.8410 & 31.03 & 0.8647 & 34.40 & 0.9291& \\
     \tr{HAT \cite{chen2023activating}} & & \tr{28.31} & \tr{0.8609} & \tr{30.21} & \tr{0.8436} & \tr{31.12} & \tr{0.8631} & \tr{34.68} & \tr{0.9291}& \\
     % KXNet & & 28.33 & 0.8627 & 30.21 & 0.8456 & 31.14 & 0.8672 & 34.59 & 0.9315 &\\
     KXNet \cite{fu2022kxnet} & & 28.51 & 0.8667 & 30.38 & 0.8485 & 31.28 & 0.8697 & 35.00 & 0.9335 &\\
     \tr{KXNet-$p4$} & & 28.91 & 0.8758 & 30.63 & 0.8564 & 31.64 & 0.8756 & 35.21 & 0.9362 &\\
     \tr{KXNet-$p8$} & & \tr{28.49} & \tr{0.8659} & \tr{30.44} & \tr{0.8505} & \tr{31.45} & \tr{0.8709} & \tr{35.08} & \tr{0.9344} &\\
     \tr{KXNet-$p8+$} & & \tr{\bf 29.13} & \tr{\bf 0.8808} & \tr{\bf 30.73} & \tr{\bf 0.8587} & \tr{\bf 31.74} & \tr{\bf 0.8772} & \tr{\bf 35.35} & \tr{\bf 0.9375} &\\
    % \midrule
    \cline{1-10}
     Bicubic & \multirow{9}{*}{x3} & 21.80 & 0.6084 & 24.68 & 0.6254 & 24.28 & 0.6546 & 25.78 & 0.7555& \\
%     RCAN \cite{zhang2018image} & & 21.38 & 0.6042 & 24.47 & 0.6299 & 24.07 & 0.6606 & 25.63 & 0.7572 &\\
     IKC \cite{gu2019blind} & & 25.36 & 0.7626 & 27.56 & 0.7475 & 28.19 & 0.7805 & 31.60 & 0.8853& \\
     DASR \cite{wang2021unsupervised} & & 25.20 & 0.7575 & 27.39 & 0.7379 & 27.96 & 0.7727 & 30.91 & 0.8723& \\
     DAN \cite{luo2020unfolding} & & 25.82 & 0.7855 & 27.88 & 0.7603 & 28.69 & 0.7969 & 31.70 & 0.8940& \\
     \tr{HAT \cite{chen2023activating}} & & \tr{26.34} & \tr{0.8026} & \tr{27.99} & \tr{0.7619} & \tr{28.78} & \tr{0.7969} & \tr{31.96} & \tr{0.8919}& \\
     % KXNet & & 26.37 & 0.8035 & 28.15 & 0.7672 & 29.04 & 0.8036 & 32.53 & 0.9034& \\
     KXNet \cite{fu2022kxnet} & & 26.48 & 0.8056 & 28.26 & 0.7699 & 29.20 & 0.8066 & 32.82 & 0.9049 &\\
     \tr{KXNet-$p4$} & & 26.66 & 0.8128 & 28.34 & 0.7742 & 29.29 & 0.8096 & 32.95 & 0.9070 &\\
     \tr{KXNet-$p8$} & & \tr{26.45} & \tr{0.8056} & \tr{28.26} & \tr{0.7709} & \tr{29.18} & \tr{0.8072} & \tr{32.75} & \tr{0.9049} &\\
     \tr{KXNet-$p8+$} & & \tr{\bf 26.95} & \tr{\bf 0.8198} & \tr{\bf 28.43} & \tr{\bf 0.7775} & \tr{\bf 29.39} & \tr{\bf 0.8124} & \tr{\bf 33.04} & \tr{\bf 0.9082} &\\
    % \midrule
    \cline{1-10}
     Bicubic & \multirow{9}{*}{x4} & 20.88 & 0.5602 & 23.75 & 0.5827 & 23.17 & 0.6082 & 24.35 & 0.7086& \\
%     RCAN \cite{zhang2018image} & & 19.84 & 0.5307 & 23.10 & 0.5729 & 22.38 & 0.5967 & 23.72 & 0.6973& \\
     IKC \cite{gu2019blind} & & 24.33 & 0.7241 & 26.49 & 0.6968 & 27.04 & 0.7398 & 29.60 & 0.8503& \\
     DASR \cite{wang2021unsupervised} & & 24.20 & 0.7150 & 26.43 & 0.6903 & 26.89 & 0.7306 & 29.53 & 0.8455& \\
     DAN \cite{luo2020unfolding} & & 24.91 & 0.7491 & 26.92 & 0.7168 & 27.69 & 0.7600 & 30.53 & 0.8746& \\
     \tr{HAT \cite{chen2023activating}} &  & \tr{24.33} & \tr{0.7231} & \tr{26.63} & \tr{0.6999} & \tr{27.18} & \tr{0.7396} & \tr{29.89} & \tr{0.8565}& \\
     % KXNet & & 25.30 & 0.7647 & 27.08 & 0.7221 & 27.98 & 0.7659 & 30.99 & 0.8815& \\
     KXNet \cite{fu2022kxnet} & & 25.41 & 0.7675 & 27.14 & 0.7233 & 28.06 & 0.7674 & 31.16 & 0.8827 &\\
     \tr{KXNet-$p4$} & & 25.55 & 0.7709 & 27.19 & 0.7240 & 28.11 & 0.7679 & 31.20 & 0.8827 &\\
     \tr{KXNet-$p8$} & & \tr{25.30} & \tr{0.7614} & \tr{27.12} & \tr{0.7202} & \tr{27.98} & \tr{0.7639} & \tr{31.08} & \tr{0.8796} &\\
     \tr{KXNet-$p8+$} & & \tr{\bf 25.69} & \tr{\bf 0.7745} & \tr{\bf 27.25} & \tr{\bf 0.7248} & \tr{\bf 28.15} & \tr{\bf 0.7682} & \tr{\bf 31.28} & \tr{\bf 0.8825} &\\
    % \bottomrule
    \hline
        %\midrule

     Bicubic & \multirow{9}{*}{x2} & 22.19 & 0.5159 & 24.44 & 0.5150 & 24.38 & 0.5497 & 25.72 & 0.6241& \multirow{27}{*}{15} \\
%     RCAN \cite{zhang2018image} & & 21.28 & 0.3884 & 22.98  & 0.3822 & 22.96 & 0.4155 & 23.76 & 0.4706& \\
     IKC \cite{gu2019blind} & & 24.69 & 0.7208 & 26.49 & 0.6828 & 26.93 & 0.7244 & 29.21 & 0.8260& \\
     DASR \cite{wang2021unsupervised} & & 24.84 & 0.7273 & 26.63 & 0.6841 & 27.22 & 0.7283 & 29.44 & 0.8322&  \\
     DAN \cite{luo2020unfolding} & & 25.32 & 0.7447 & 26.84 & 0.6932 & 27.56 & 0.7392 & 29.91 & 0.8430& \\
     \tr{HAT \cite{chen2023activating}} & & \tr{25.60} & \tr{0.7550} & \tr{26.91} & \tr{0.6962} & \tr{27.69} & \tr{0.7429} & \tr{30.02} & \tr{0.8452}& \\
     % KXNet & & 25.45 & 0.7500 & 26.87 & 0.6959 & 27.59 & 0.7422 & 29.93 & 0.8449& \\
     KXNet \cite{fu2022kxnet} & & 25.45 & 0.7492 & 26.87 & 0.6943 & 27.61 & 0.7409 & 29.95 & 0.8437 &\\
     \tr{KXNet-$p4$} & & 25.62 & 0.7578 & 26.91 & 0.6980 & 27.69 & 0.7435 & 30.00 & 0.8452 &\\
     \tr{KXNet-$p8$} & & \tr{25.48} & \tr{0.7508} & \tr{26.89} & \tr{0.6957} & \tr{27.65} & \tr{0.7420} & \tr{29.96} & \tr{0.8435} &\\
     \tr{KXNet-$p8+$}& & \tr{\bf 25.73} & \tr{\bf 0.7613} & \tr{\bf 26.94} & \tr{\bf 0.6985} & \tr{\bf 27.75} & \tr{\bf 0.7450} & \tr{\bf 30.06} & \tr{\bf 0.8463} &\\
    % \midrule
    % \midrule
    \cline{1-10}
     Bicubic & \multirow{9}{*}{x3} & 21.18 & 0.4891 & 23.55 & 0.4961 & 23.28 & 0.5289 & 24.42 & 0.6119& \\
%     RCAN \cite{zhang2018image} & & 20.22 & 0.3693 & 22.20 & 0.3726 & 21.99 & 0.4053 & 22.85 & 0.4745& \\
     IKC \cite{gu2019blind} & & 24.21 & 0.7019 & 25.93 & 0.6564 & 26.42 & 0.7018 & 28.61 & 0.8135& \\
     DASR \cite{wang2021unsupervised} & & 23.93 & 0.6890 & 25.82 & 0.6484 & 26.27 & 0.6940 & 28.27 & 0.8047& \\
     DAN \cite{luo2020unfolding} & & 24.17 & 0.7013 & 25.93 & 0.6551 & 26.46 & 0.7014 & 28.52 & 0.8130& \\
     % KXNet & & 24.42 & 0.7135 & 25.99 & 0.6585 & 26.56 & 0.7063 & 28.64 & 0.8178& \\
     \tr{HAT \cite{chen2023activating}} & & \tr{24.57} & \tr{0.7171} & \tr{26.01} & \tr{0.6574} & \tr{26.58} & \tr{0.7053} & \tr{28.64} & \tr{0.8150}& \\
     KXNet \cite{fu2022kxnet} & & 24.42 & 0.7117 & 26.00 & 0.6574 & 26.60 & 0.7054 & 28.69 & 0.8162 &\\
     \tr{KXNet-$p4$} & & 24.53 & 0.7181 & 26.03 & 0.6595 & 26.62 & 0.7074 & 28.75 & 0.8191 &\\
     \tr{KXNet-$p8$} & & \tr{24.44} & \tr{0.7133} & \tr{26.00} & \tr{0.6571} & \tr{26.61} & \tr{0.7054} & \tr{28.74} & \tr{0.8174} &\\
     \tr{KXNet-$p8+$} & & \tr{\bf 24.68} & \tr{\bf 0.7220} & \tr{\bf 26.05} & \tr{\bf 0.6598} & \tr{\bf 26.68} & \tr{\bf 0.7089} & \tr{\bf 28.79} & \tr{\bf 0.8199} &\\
    % \midrule
    \cline{1-10}
     Bicubic & \multirow{9}{*}{x4} & 20.38 & 0.4690 & 22.83 & 0.4841 & 22.39 & 0.5120 & 23.33 & 0.5977& \\
%     RCAN \cite{zhang2018image} & & 19.23 & 0.3515 & 21.47 & 0.3686 & 21.05 & 0.3960 & 21.77 & 0.4689& \\
     IKC \cite{gu2019blind} & & 23.35 & 0.6665 & 25.21 & 0.6238 & 25.58 & 0.6712 & 27.45 & 0.7867& \\
     DASR \cite{wang2021unsupervised} & & 23.26 & 0.6620 & 25.20 & 0.6223 & 25.55 & 0.6683 & 27.32 & 0.7842& \\
     DAN \cite{luo2020unfolding} & & 23.48 & 0.6742 & 25.25 & 0.6283 & 25.72 & 0.6760 & 27.55 & 0.7938& \\
     \tr{HAT \cite{chen2023activating}} & & \tr{23.27} & \tr{0.6634} & \tr{25.20} & \tr{0.6232} & \tr{25.57} & \tr{0.6697} & \tr{27.47} & \tr{0.7883} & \\
     % KXNet & & 23.67 & 0.6844 & 25.30 & 0.6296 & 25.78 & 0.6792 & 27.66 & 0.7977&\\
     KXNet \cite{fu2022kxnet} & & 23.70 & 0.6842 & 25.30 & 0.6293 & 25.82 & 0.6790 & 27.71 & 0.7979 &\\
     \tr{KXNet-$p4$} & & 23.84 & 0.6895 & 25.34 & \bf 0.6306 & 25.88 & 0.6805 & 27.79 & 0.7996 &\\
     \tr{KXNet-$p8$} & & \tr{23.75} & \tr{0.6853} & \tr{25.33} & \tr{0.6297} & \tr{25.85} & \tr{0.6788} & \tr{27.75} & \tr{0.7980} &\\
     \tr{KXNet-$p8+$} & & \tr{\bf 23.91} & \tr{\bf 0.6915} & \tr{\bf 25.35} & \tr{0.6300} & \tr{\bf 25.90} & \tr{\bf 0.6808} & \tr{\bf 27.86} & \tr{\bf 0.8001} &\\
    \bottomrule
    % \hline
  \end{tabular}
%   \vspace{4mm}
  \label{tab:table_sr}
  \vspace{-3mm}
\end{table*}
%-----------------------

\section{Experimental results}

\subsection{Blind Single Image Super-Resolution}\label{KXNet_exp}

\begin{figure*}[ht]
\centering
% \vspace{1mm}
\subfigure{
    \begin{minipage}[b]{0.185\linewidth}
    \centerline{\includegraphics[width=3.5cm]{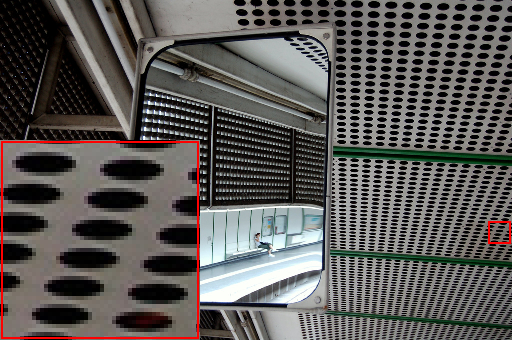}}
    \vspace{-0.8mm}
    \centerline{GT}
    \centerline{PSNR/SSIM}
    \end{minipage}
    } %[图片大小]{图片路径}
\subfigure{
    \begin{minipage}[b]{0.185\linewidth}
    \centerline{\includegraphics[width=3.5cm]{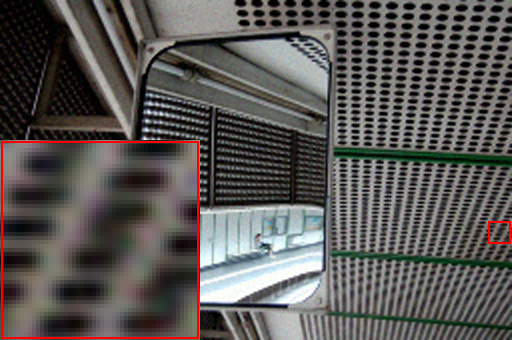}}
    \vspace{-0.8mm}
    \centerline{Bicubic} %\vspace{0.8mm}
    \centerline{18.65/0.5466}
    \end{minipage}
    } %[图片大小]{图片路径}
\subfigure{
    \begin{minipage}[b]{0.185\linewidth}
    \centerline{\includegraphics[width=3.5cm]{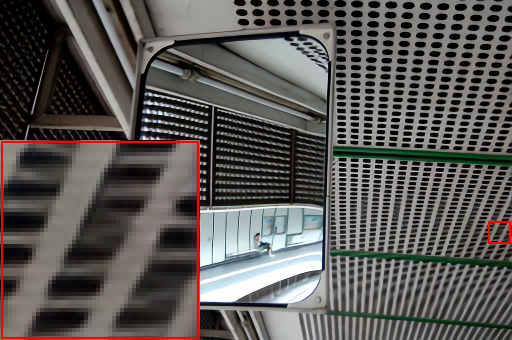}}
    \vspace{-0.8mm}
    \centerline{IKC}
    \centerline{22.19/0.7942}
    \end{minipage}
    } %[图片大小]{图片路径}
\subfigure{
    \begin{minipage}[b]{0.185\linewidth}
    \centerline{\includegraphics[width=3.5cm]{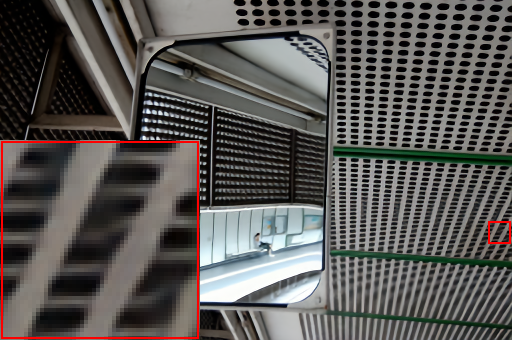}}
    \vspace{-0.8mm}
    \centerline{DASR}
    \centerline{22.06/0.7761}
    \end{minipage}
    } %[图片大小]{图片路径}
\subfigure{
    \begin{minipage}[b]{0.185\linewidth}
    \centerline{\includegraphics[width=3.5cm]{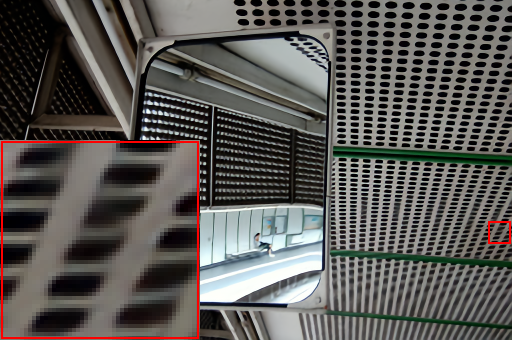}}
    \vspace{-0.8mm}
    \centerline{DAN} 
    \centerline{22.08/0.7939}
    \end{minipage}
    } %[图片大小]{图片路径}
\subfigure{
    \begin{minipage}[b]{0.185\linewidth}
    \centerline{\includegraphics[width=3.5cm]{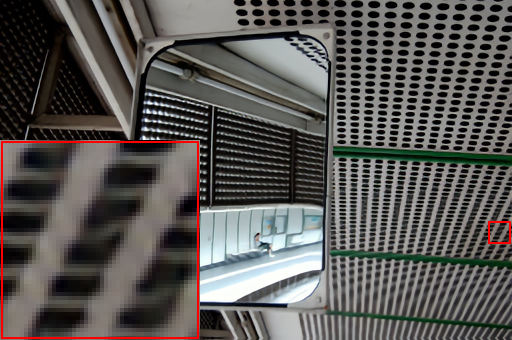}}
    \vspace{-0.8mm}
    \centerline{HAT} 
    \centerline{21.99/0.7743}
    \end{minipage}
    } %[图片大小]{图片路径}
\subfigure{
    \begin{minipage}[b]{0.185\linewidth}
    \centerline{\includegraphics[width=3.5cm]{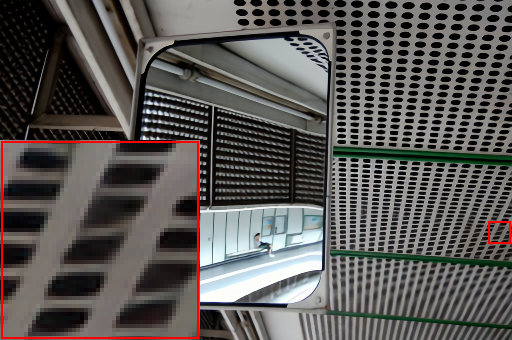}}
    \vspace{-0.8mm}
    \centerline{KXNet}
    \centerline{22.31/0.8046}
    \end{minipage}
    } %[图片大小]{图片路径}
\subfigure{
    \begin{minipage}[b]{0.185\linewidth}
    \centerline{\includegraphics[width=3.5cm]{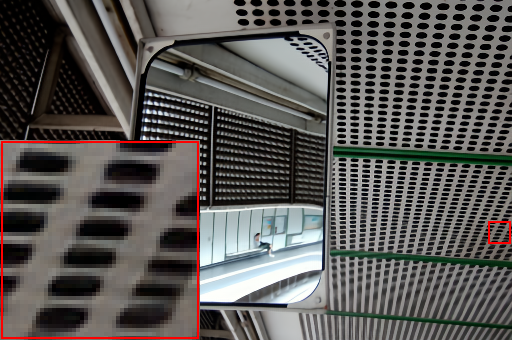}}
    \vspace{-0.8mm}
    \centerline{KXNet-$p4$}
    \centerline{22.62/0.8099}
    \end{minipage}
    } %[图片大小]{图片路径}
\subfigure{
    \begin{minipage}[b]{0.185\linewidth}
    \centerline{\includegraphics[width=3.5cm]{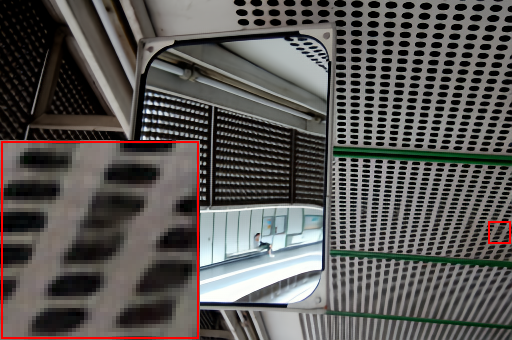}}
    \vspace{-0.8mm}
    \centerline{KXNet-$p8$} 
    \centerline{22.29/0.7984}
    \end{minipage}
    } %[图片大小]{图片路径}
\subfigure{
    \begin{minipage}[b]{0.185\linewidth}
    \centerline{\includegraphics[width=3.5cm]{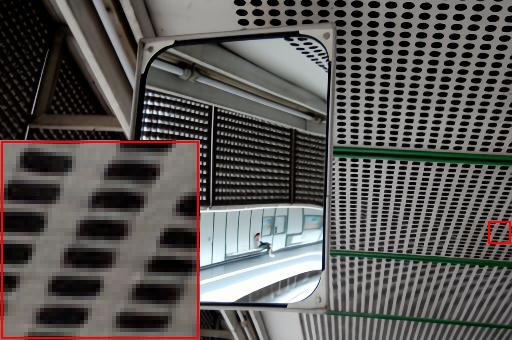}}
    \vspace{-0.8mm}
    \centerline{KXNet-$p8+$} 
    \centerline{\textbf{22.87}/\textbf{0.8167}}
    \end{minipage}
    } %[图片大小]{图片路径}
% \subfigure[Jackson Yee]{\includegraphics[width=3.5cm]{image/014GT.png}}
% \subfigure[Jackson Yee]{\includegraphics[width=3.5cm]{image/014GT.png}}
\vspace{-5mm}
\caption{Performance comparison on \textit{img 004} in Urban100 \cite{huang2015single}. The scale factor is 4 and the standard deviation is 5.} %图片标题
\label{fig:figure3}  %图片交叉引用时的标签
\vspace{-1mm}
\end{figure*}
%-----------------------------------------

We first test the effectiveness of the equivariant proximal work using the blind image super-resolution task.
For image super-resolution, the main goal is to reconstruct the high-resolution image with high visual quality from an observed low-resolution image. The degradation model for blind super-resolution with a uniform kernel is commonly expressed as:
\begin{equation}
    {Y} = ({X}\otimes {K}) \downarrow_{{s}} + {N},
    \label{eq:eq5}
\end{equation}
where ${X}$ is the to-be-estimated high-resolution image, ${K}$ is the blur kernel, $\otimes$ denotes two-dimensional (2D) convolution operation,  $\downarrow _{{{s}}}$ represents the standard $s$-fold downsampler, \textit{i.e.}, only keeping the upper-left pixel for each distinct ${s\times s}$ patch \cite{zhang2020deep}, and ${N}$ denotes the Additive White Gaussian Noise (AWGN).

Estimating ${X}$ and ${K}$ from ${Y}$ constitutes a challenging ill-posed inverse problem. Nevertheless, the individual subproblems pertaining to the estimation of ${X}$ and ${K}$ can be classified as instances of the broader inverse problem (\ref{eq:eq1}). By iteratively adapting the proximal gradient descent algorithm and deep unfolding technique  for solving $K$ and $X$, in a manner similar to Eqs. (\ref{eq:ista})-(\ref{eq:ista_prox}), \cite{fu2022kxnet} proposed the SOTA deep unfolding network, namely, KXNet.
Specifically, the update formula for $K$ is
\begin{equation}
   {{K}}^{(t+1)} = \operatorname{prox}_{\theta_{ K}^{(t)}} \left( {{K}}^{(t)} - \delta_1 \nabla f \left( {{K}}^{(t)} \right) \right),
    \label{eq:eq7}
\end{equation}
where $\operatorname{proxNet}_{\theta_{K}^{(t)}} (\cdot)$ is a shallow ResNet with the parameters $\theta_{ K}^{(t)}$. Due to space limit, we don't represent the specific form of $\nabla f ({{K}}^{\left(t-1\right)})$ here, which can be referred to \cite{fu2022kxnet}.
Besides, the update formula for $ {X}$ is:
\begin{equation}
% \small
% \hspace{-0.3mm}
    \begin{split}
    & {X}^{(t+1)} = \\
    &\!\!\operatorname{prox}_{\theta_{ X}^{(t)}}\! \!\l( {X}^{(t)} \!\!-\!
     \delta_2 {K}^{(t+1)} \!\!\otimes_{{s}}^{\mathrm{T}}\! \left( {Y} \!\!-\! \l({X}^{(t)} \!\otimes\! {K}^{(t+1)}\r)\! \downarrow_{{s}} \right)\!\r),
    \end{split}
    \label{eq:eq9}
\end{equation}
where $\operatorname{proxNet}_{\theta_{ X}^{(t)}} (\cdot)$ is a shallow ResNet with the parameters $\theta_{ X}^{(t)}$, and $\otimes_{{s}}^{\mathrm{T}}$ denotes the transposed convolution operation with stride as ${s}$.

\begin{table*}[t]
  \centering % \setlength{\tabcolsep}{40pt}
    \caption{Generalization results of all comparison methods averaged on 50 standard deviation, while the model is trained with 0-25 standard deviation.}
    \centering \setlength{\tabcolsep}{14pt}
\begin{tabular}{l c c c c c c c c c c}
    \toprule
    % \hline
     \multirow{2}{*}{Method} & \multirow{2}{*}{Scale} & \multicolumn{2}{c}{Urban100 \cite{huang2015single}} & \multicolumn{2}{c}{BSD100 \cite{martin2001database}} & \multicolumn{2}{c}{Set14 \cite{zeyde2010single}} & \multicolumn{2}{c}{Set5 \cite{bevilacqua2012low}} \\
 \cmidrule(r){3-4}  \cmidrule(r){5-6} \cmidrule(r){7-8} \cmidrule(r){9-10}
     & & PSNR & SSIM & PSNR & SSIM & PSNR & SSIM & PSNR & SSIM \\
    % \midrule
    \midrule
    % \hline
    % \hline
%      \tr{HAT} & \multirow{4}{*}{x2} & \tr{23.27} & \tr{0.6540} & \tr{24.76} & \tr{0.6017} & \tr{25.17} & \tr{0.6520} & \tr{26.69} & \tr{0.7645} \\
      KXNet & \multirow{4}{*}{x2} & 23.27 & 0.6586 & 24.80 & 0.6071 & 25.14 & 0.6532 & 26.62 & 0.7652 \\
      \tr{KXNet-$p4$} & & 23.40 & 0.6653 & 24.84 & 0.6079 & 25.18 & 0.6541 & 26.66 & 0.7678 \\
      \tr{KXNet-$p8$} & & \tr{23.35} & \tr{0.6610} & \tr{24.85} & \tr{0.6074} & \tr{25.19} & \tr{0.6548} & \tr{\bf 26.75} & \tr{0.7692} \\
      \tr{KXNet-$p8+$} & & \tr{\bf 23.55} & \tr{\bf 0.6727} & \tr{\bf 24.89} & \tr{\bf 0.6102} & \tr{\bf 25.29} & \tr{\bf 0.6586} & \tr{26.74} & \tr{\bf 0.7715} \\
    % \midrule
    \cline{1-10} \vspace{-2mm}\\
%     \tr{HAT} & \multirow{4}{*}{x3} & \tr{22.44} & \tr{0.6209} & \tr{23.97} & \tr{0.5747} & \tr{24.14} & \tr{0.6181} & \tr{25.27} & \tr{0.7216} \\
     KXNet & \multirow{4}{*}{x3}& 22.28 & 0.6162 & 23.86 & 0.5742 & 24.08 & 0.6172 & 25.23 & 0.7229 \\
     \tr{KXNet-$p4$} & & 22.54 & 0.6307 & 24.06 & 0.5782 & 24.26 & 0.6229 & 25.43 & 0.7330\\
     \tr{KXNet-$p8$} & & \tr{22.51} & \tr{0.6262} & \tr{24.06} & \tr{0.5765} & \tr{24.25} & \tr{0.6211} & \tr{25.50} & \tr{0.7331} \\
     \tr{KXNet-$p8+$} & & \tr{\bf 22.70} & \tr{\bf 0.6357} & \tr{\bf 24.09} & \tr{\bf 0.5791} & \tr{\bf 24.34} & \tr{\bf 0.6253} & \tr{\bf 25.54} & \tr{\bf 0.7364} \\
    \cline{1-10} \vspace{-2mm} \\
%     \tr{HAT} & \multirow{4}{*}{x4} & \tr{21.22} & \tr{0.5546} & \tr{22.91} & \tr{0.5244} & \tr{22.89} & \tr{0.5587} & \tr{23.74} & \tr{0.6553} \\
     KXNet & \multirow{4}{*}{x4}  & 21.83 & 0.5978 & 23.34 & 0.5532 & 23.45 & 0.5954 & 24.42 & 0.7047 \\
     \tr{KXNet-$p4$} & & 21.99 & 0.6064 & 23.49 & 0.5575 & 23.58 & 0.5999 & 24.55 & 0.7094 \\
     \tr{KXNet-$p8$} & & \tr{21.96} & \tr{0.6035} & \tr{23.48} & \tr{0.5566} & \tr{23.55} & \tr{0.5981} & \tr{24.55} & \tr{0.7066} \\
     \tr{KXNet-$p8+$} & & \tr{\bf 22.06} & \tr{\bf 0.6082} & \tr{\bf 23.50} & \tr{\bf 0.5575} & \tr{\bf 23.60} & \tr{\bf 0.6009} & \tr{\bf 24.60} & \tr{\bf 0.7105}\\
    \bottomrule
    % \hline
  \end{tabular}
%   \vspace{4mm}
  \label{tab:sr_gener}
   \vspace{-2mm}
\end{table*}

\tr{The original KXNet  employs regular convolution for constructing $\operatorname{proxNet}_{\theta_x^{(t)}} (\cdot)$ and updating $X$, which neglects the rotation symmetry prior in an image. We thus exploit rotation equivariant convolution, F-Conv \cite{xie2022fourier}, to replace the regular convolution and construct a rotation equivariant proximal network for updating $X$. 
}

{\bf {Network architecture Setting.}}
\tr{In order to  comprehensively verify the effects of rotational equivariance in the proximal network, we construct the equivariant proximal network under 3 different network architecture settings and test their performance. The first one is named KXNet-$p4$, where the equivariant number $t$ is set as 4, i.e., the network is rotation equivariant with $\sfrac{\pi}{2}$ rotations. In this network, the channel number of rotation equivariant convolution is set as $\sfrac{1}{4}$ to the original regular convolution based network, such that the memory usage is similar to the original network. The second one is named KXNet-$p8$ which is with equivariant number $t = 8$, and  channel number is set as $\sfrac{1}{8}$ to the  regular convolution based network.  Note that the parameter in  KXNet-$p8$ is much fewer than KXNet-$p4$ and the original KXNet, since the parameter number in each convolution layer is only  $\sfrac{1}{8}$  to the regular convolution layer. Thus,  We  further construct  KXNet-$p8$+, which is also with $t = 8$, but double the channel number to make its parameter number similar to KXNet-$p4$. Besides, we replace $3\times 3$ size filter in the original network with the $5\times 5$ filter in F-Conv for better rotating the filters.}
The training settings and loss function follow the original settings. All the network parameters can be automatically learned from training data in an end-to-end manner.

% \subsection{Datasets and Training settings}
{\bf Datasets and Training settings.}
Following \cite{fu2022kxnet}, we use 3450 training images from DIV2K \cite{agustsson2017ntire} and Flickr2K \cite{timofte2017ntire} datasets. For testing, we adopt commonly-used benchmark datasets, including Set5 \cite{bevilacqua2012low}, Set14 \cite{zeyde2010single}, BSD100 \cite{martin2001database} and Urban100 \cite{huang2015single}. We conduct experiments with anisotropic Gaussian blur kernel and Gaussian noise. For the training set, we set the kernel size $p$ as $11/15/21$ for $\times 2/3/4$ SR, respectively. The kernel width at each axis is obtained by randomly rotating the widths $\lambda_1$ and $\lambda_2$ with an angle $\theta \sim U[-\pi, \pi]$, where $\lambda_1$ and $\lambda_2$ are uniformly distributed in $U(0.6, 5.0)$. Besides, the range of standard deviation $\sigma$ for Gaussian noise is set as $[0, 25]$. For the testing datasets, we separately set the kernel width as $\lambda_1=0.8, \lambda_2=1.6$ and $\lambda_1=2.0, \lambda_2=4.0$, and rotate them by $\theta\in\{0, \sfrac{\pi}{4}, \sfrac{\pi}{2}, \sfrac{3 \pi}{4}\}$, respectively. This means every HR testing image is degraded by 8 different blur kernels.

% \subsection{Quantitative and Qualitative Comparison}
{\bf Quantitative and qualitative comparison.}
\tr{As shown in Table~\ref{tab:table_sr}, the KXNet-$p4$, KXNet-$p8$, KXNet-$p8$+ consistently demonstrate superior performance compared to other competing methods across four benchmark datasets 
that encompass  varying SR scales and standard deviations, while KXNet-$p4$ performs slightly better than KXNet-$p8$, and KXNet-$p8$+ achieves the best performance. 
} Fig. \ref{fig:figure3} visually shows the SR results of all competing methods on one typical sample. The proposed method clearly yields better restoration results, particularly in regions containing recurrent rotation symmetry structures within the image.

\begin{figure}
    \centering
    \includegraphics[width=\linewidth, scale=1.0]{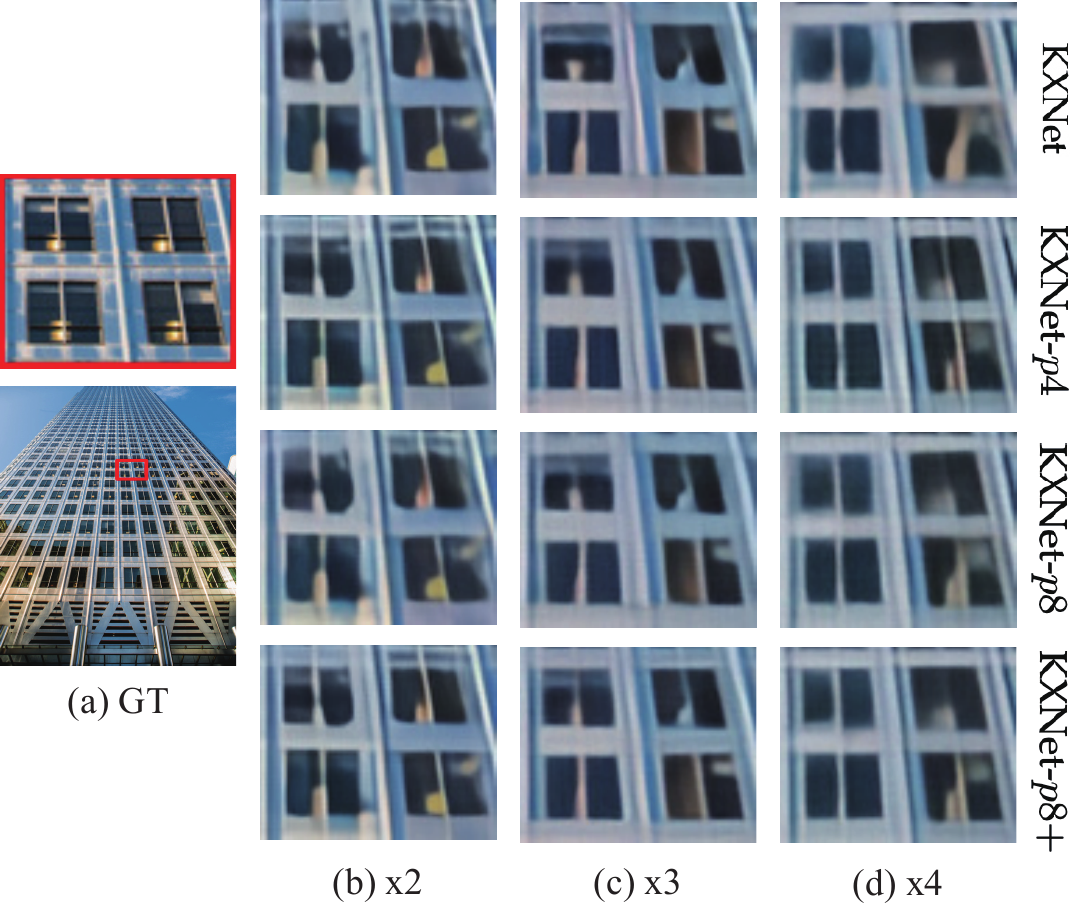}
    \caption{Performance comparison on \textit{img 030} in Urban100 \cite{huang2015single}. The standard deviation is 50.} %图片标题
    \label{fig:sr_generlization}  %图片交叉引用时的标签
    \vspace{-4mm}
\end{figure}

{\bf Generalization performance.}
To verify the generalization capability of the rotation equivariant proximal network, we further test the competing methods in new scenarios different from the training set. In Table.~\ref{tab:sr_gener}, we evaluate the comparison methods with AWGN whose standard deviation is 50, whereas the standard deviation in the training set ranges from 0 to 25. It can be observed that the image reconstruction accuracy of the enhanced KXNet significantly outperforms the original KXNet.  Fig. \ref{fig:sr_generlization} is the corresponding visualization result. The superiority of the rotation equivariant amelioration is also evident.

\begin{figure}
    \centering
    \includegraphics[width=\linewidth, scale=1.0]{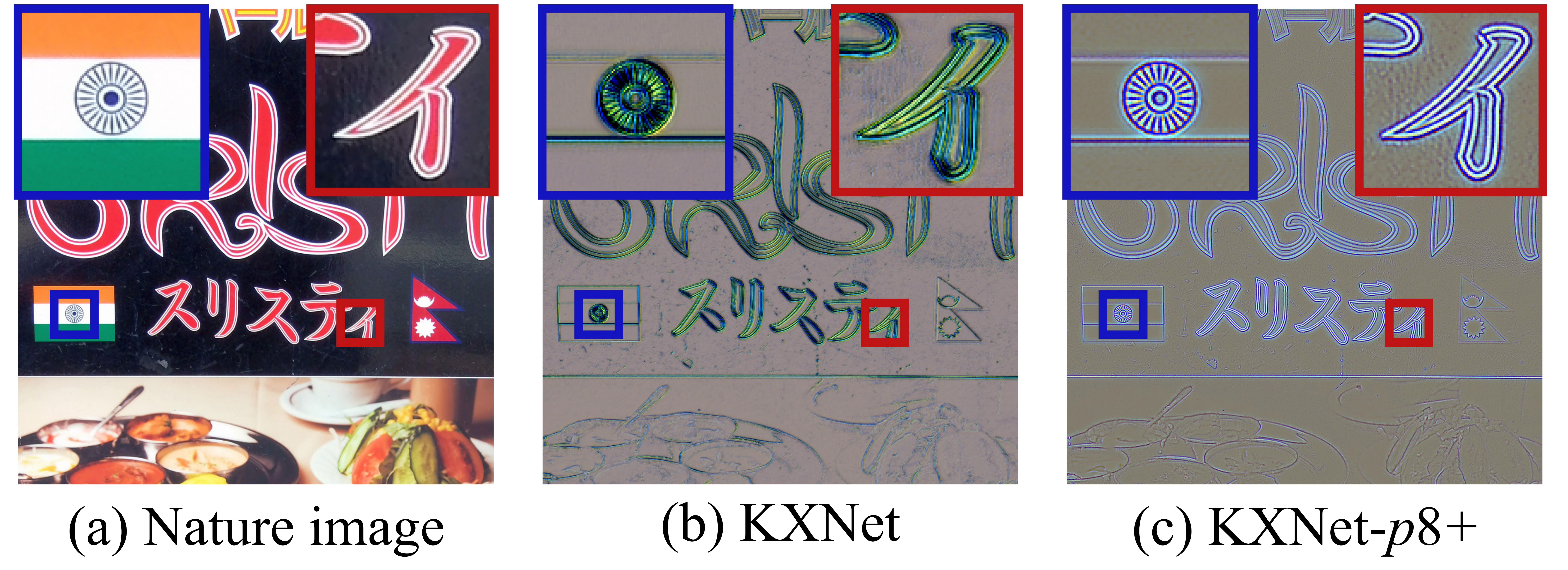}
    \vspace{-5mm}
    \caption{\tr{The feature maps of the KXNet and KXNet-$p8$+ on \textit{img 000096} in Flickr2K. The scale factor is 2.}}
    \label{fig:feature_map_sr}
    \vspace{-1mm}
\end{figure}

\tr{
{\bf Feature maps visualization.}
To validate the rotation equivariance of the proposed method, we conduct visualization of feature maps generated by the proximal network of  KXNet and KXNet-$p8+$, respectively. As illustrated in Fig. \ref{fig:feature_map_sr}, the feature map of the proposed method shows superior rotational symmetry. One can refer to the supplementary material for more equivariance validation demonstrations.}

\subsection{Metal Artifact Reduction in CT Image}\label{CT_exp}
During the computed tomography (CT) imaging process, metallic implants within patients would severely attenuate X-ray and even lead to missing X-ray projections. Accordingly, the captured CT images often present streaking and shading artifacts \cite{liao2019adn} \cite{lin2019dudonet}, which negatively affect the clinical treatment. Therefore, the task of metal artifact reduction (MAR) in CT images has received extensive attention in the medical imaging field.

\begin{figure}
    \centering
    \includegraphics[width=\linewidth, scale=1.0]{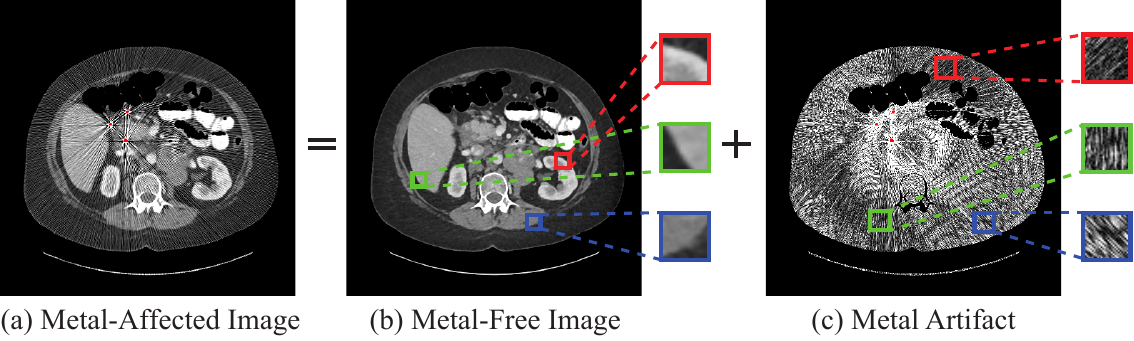}
    \caption{(a) A metal-affected CT image of human tissue structures. (b) The metal-free part contains rotation symmetry visual patterns. (c) The metal artifacts part contains rotationally symmetrical streaking prior structures.}
    \label{fig:mar01}    \vspace{-1mm}
\end{figure}

Normally, an observed metal-affected CT image consists of two parts, i.e. metal-free part and metal artifact part, as illustrated in Fig. \ref{fig:mar01}.
%Obviously, the metals generally have extremely higher CT values than normal tissues.
%Thus, for the CT metal artifact reduction task, our goal is to remove unexpected metal artifacts and restore normal tissue structure in non-metal regions.
In the SOTA method \cite{wang2021dicdnet} for metal artifacts removing,
the degradation model for a metal-corrupted CT image ${Y} \in \mathbb{R}^{H\times W}$, is modeled as the following expression:
\begin{equation}
    {I} \odot {Y} = {I} \odot {X} + {I} \odot {M}
    \label{eq:eq12}
\end{equation}
where $ {X},{M} \in \mathbb{R}^{H \times W}$ are the to-be-estimated clean metal-free CT image and the metal artifact layer, respectively, ${I} \in \mathbb{R}^{H \times W}$ is a binary mask representing the non-metal region, and $\odot$ is the point-wise multiplication operation.

%----------------------------------
%-------------------------------------

%----------------------------------
%-------------------------------------

 %-------------------------------------------
\begin{figure*}[t]
    \centering
    \includegraphics[width=18.2cm]{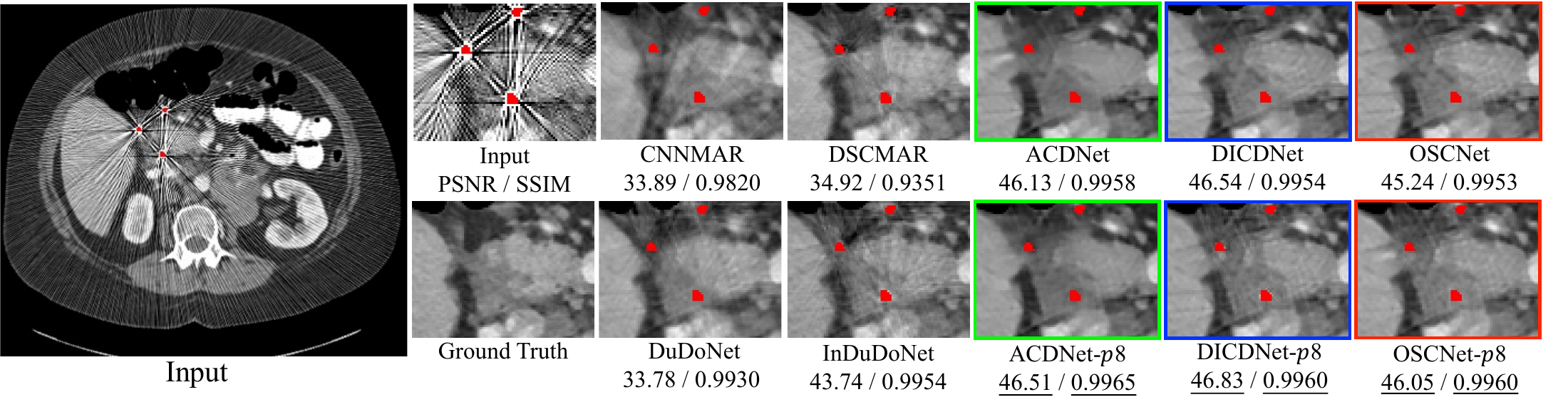}\vspace{-2mm}
    \caption{Performance comparison on a typical
metal-corrupted CT image from the synthesized DeepLesion \cite{yan2018deep}. The red pixels stand for metallic implants.}
    \label{fig:mar_deeplesion}\vspace{-2mm}
\end{figure*}
%---------------------------------------------

Combining the model (\ref{eq:eq12}) with different metal artifact models, multiple algorithms and unfolding networks for metal artifact removal have been proposed.
In the following, we adopt the suggested rotation equivariant proximal network into three SOTA unfolding networks for the task.
%In DICDNet\cite{wang2021dicdnet},

{\bf Convolutional dictionary network for MAR.}
Wang et al. proposed DICDNet\cite{wang2021dicdnet}, introducing a convolution dictionary mechanism to encode the structural prior knowledge of metal artifact layer $ M$,
\begin{equation}
    {M} = \sum\limits_{n=1}^{N} {K}_n \otimes {M}_n = \mathcal{K} \otimes \mathcal{M},
    \label{eq:eq13}
\end{equation}
where ${K}_n \in \mathbb{R}^{f\times f}$ is a convolutional kernel with size $f\times f$, denoting the local patterns of artifacts, and ${M}_n \in \mathbb{R}^{H \times W}$ is the corresponding feature map denoting the locations where local patterns appear, $N$ is the number of kernels and $\otimes$ is the 2D convolutional operation. For convenience, convolution and summation are further written in tensor form, i.e. $\mathcal{K} \otimes \mathcal{M}$.

By substituting (\ref{eq:eq13}) into (\ref{eq:eq12}),  the subproblems for estimating $\mathcal{M}$ and ${X}$  are instances of the general inverse problem (\ref{eq:eq1}).
Then by manners similar as Eqs. (\ref{eq:ista})-(\ref{eq:ista_prox}), DICDNet\cite{wang2021dicdnet}  got the network iterative structure of $\mathcal{M}$ and $ X$ as:
% \begin{equation}
% \begin{split}
%     &\mathcal{M}^{(t)} = \\
%     & \operatorname{prox}_{\alpha \eta_1} ( \mathcal{M}^{(t-1)} - \eta_1 \mathcal{K} \otimes^{T} {I} \odot ( \mathcal{K}\otimes \mathcal{M}^{(s-1)} + {X}^{(s-1)} - {Y} )))
% \end{split}
%     \label{eq:eq15}
% \end{equation}
%%%%%%%%%%%%%%%%%%%%%%%%%%%%%%%%%%%%%%%%%%%%%%%%
\begin{equation}
    \left \{
    \begin{aligned}
         \mathcal{M}^{(t)} & = \operatorname{prox}_{\theta^{(t)}_{\mathcal{M}}}(\mathcal{M}^{(t-1)}-\eta_1 \nabla g_1(\mathcal{M}^{(t-1)})), \\
         {X}^{(t)} & = \operatorname{prox}_{\theta^{(t)}_{{X}}} ({X}^{(t-1)} - \eta_2 \nabla g_2({X}^{(t-1)})),
    \end{aligned}
    \right.
    \label{eq:eq15}
\end{equation}
where $\nabla g_1(\mathcal{M}^{(t-1)}) = \mathcal{K} \otimes^{T} ({I} \odot ( \mathcal{K}\otimes \mathcal{M}^{(s-1)} + {X}^{(s-1)} - {Y} ))$, $\nabla g_2({X}^{(t-1)}) = {I} \odot (\mathcal{K} \otimes \mathcal{M}^{(n)} + {X}^{(t-1)} - {Y})$, $\otimes ^{T}$ denotes the transposed convolution operation, $\operatorname{prox}_{\theta^{(t)}_{\mathcal{M}}}(\cdot)$ and $\operatorname{prox}_{\theta^{(t)}_{{X}}}(\cdot)$ are proximal operator networks consisting of a simple ResNet structure, respectively. Among these operations, $\otimes ^{T}$, $\otimes$, $\odot$ can be easily implemented in Pytorch.

{\bf Orientation-shared convolution for MAR.}
% \noindent {\bf Orientation-Shared Convolution Representation for MAR}
Since the acquisition process of CT images is generally performed in a rotationally scanning manner, metal artifacts generally exhibit scattered streak structures,  i.e., rotationally symmetrical streaking (RSS) prior {as shown in}
Fig. \ref{fig:mar01} (c).  In order to encode such inherent RSS prior structures, OSCNet \cite{wang2022orientation} proposed a convolutional dictionary model based on filter parameterization that shares convolution filters among different angles:
\begin{equation}
    {M} = \sum\limits_{l=1}^{L} \sum\limits_{k=1}^{K} {C}_{k}(\theta_{l}) \otimes {M}_{lk} = \mathcal{C} \otimes \mathcal{M},
    \label{eq:eq16}
\end{equation}
where $L$ is the number of rotation angles, $\theta_l = \sfrac{2 \pi (l-1)}{L}$ is the $l^{th}$ rotation angles, $K$ is the number of convolution filters at each angle, ${C}_{k}(\theta_{l}) \in \mathbb{R}^{p \times p}$ is the $k^{th}$ parameterized filter at angle $\theta_l$, and it represents the streaking and rotated prior patterns of artifacts, ${M}_{lk}$ is feature map marking the location of artifacts, and $\mathcal{C}\!\in\! \mathbb{R}^{p\times p \times LK}$ and $\mathcal{M} \!\in\! \mathbb{R}^{H \times W \times LK}$ are stacked by ${C}_k(\theta_l)$ and ${M}_{lk}$, respectively.

Similar to (\ref{eq:eq15}), the following iteration updating formulas of $X$ and $\mathcal{M}$ for model (\ref{eq:eq16}) can be deduced as \cite{wang2022orientation}:
\begin{equation}
    \left \{
    \begin{aligned}
         \mathcal{M}^{(t)} & = \operatorname{prox}_{\theta^{(t)}_{\mathcal{M}}}(\mathcal{M}^{(t-1)}-\eta_1 \nabla g_1(\mathcal{M}^{(t-1)})), \\
         {X}^{(t)} & = \operatorname{prox}_{\theta^{(t)}_{{X}}} ({X}^{(t-1)} - \eta_2 \nabla g_2({X}^{(t-1)})),
    \end{aligned}
    \right.
    \label{eq:eq18}
\end{equation}
where $\nabla g_1(\mathcal{M}^{(t-1)}) = \mathcal{C} \otimes^{T} ({I} \odot ( \mathcal{C}\otimes \mathcal{M}^{(s-1)} + {X}^{(s-1)} - {Y} ))$, $\nabla g_2({X}^{(t-1)}) = {I} \odot (\mathcal{C} \otimes \mathcal{M}^{(n)} + {X}^{(t-1)} - {Y})$, $\otimes ^{T}$ denotes the transposed convolution operation, $\operatorname{prox}_{\theta^{(t)}_{\mathcal{M}}}(\cdot)$ and $\operatorname{prox}_{\theta^{(t)}_{{X}}}(\cdot)$ are proximal  networks consisting of a simple ResNet structure, respectively.

{\bf Adaptive Convolution Dictionary for MAR.}
% \noindent {\bf Adaptive Convolution Dictionary for MAR}
In order to  bridge the domain gap between the real scenarios and training data, an adaptive dictionary learning method for MAR is proposed, namely ACDNet \cite{wang2022adaptive}. Specifically, a metal artifact map $ M$ is obtained through a weighted convolutional dictionary model,
\begin{equation}
    {M} = \sum\limits_{n=1}^{N} (\mathcal{D} *  W_n) \otimes  M_n = (\mathcal{D} *  W) \otimes \mathcal{M},
    \label{eq:eqacdA}
\end{equation}
where $\mathcal{D}\in \mathbb{R}^{p\times p \times d}$ is a sample-invariant dictionary representing common local patterns of metal artifacts in CT images, $W_n \in \mathbb{R}^{d}$ denotes weighted coefficient, and $\mathcal{D}*  W \otimes \mathcal{M}$ denotes the tensor form for convenience.

Similar to (\ref{eq:eq15}) and (\ref{eq:eq18}), the network iteration process for $ W$, $\mathcal{M}$ and $ X$ can be then deduced, which is with the following formulation \cite{wang2022adaptive}:
\begin{equation}
    \left \{
    \begin{aligned}
         W^{(t)} & = \operatorname{prox}_{\theta_{ W}^{(t)}} \left(  W^{(t-1)} - \eta_1 \nabla g_1 \left( W^{(t-1)} \right) \right) \\
        \mathcal{M}^{(t)} & = \operatorname{prox}_{\theta_{\mathcal{M}}^{(t)}} \left( \mathcal{M}^{(t-1)} - \eta_2 \nabla g_2 \left(\mathcal{M}^{(t-1)} \right) \right) \\
         X^{(t)} & = \operatorname{prox}_{\theta_{ X}^{(t)}} \left(  X^{(t-1)} - \eta_3 \nabla g_3 \left( X^{(t-1)} \right) \right),
    \end{aligned}
    \right.
    \label{eq:eqacdIter}
\end{equation}
where $\nabla g_1 ( W^{(t-1)}) = \sfrac{\partial g_1( W^{(t-1)})}{\partial  W}$ ,$\nabla g_2 (\mathcal{M}^{(t-1)}) = (\mathcal{D}* {W}^{(t)}) \otimes^{T} ( I \odot ((\mathcal{D}*{W}^{(t)}) \otimes \mathcal{M}^{(t-1)} + {X}^{(t-1)} -  Y))$, $\nabla g_3({X}^{(t-1)}) =  I \odot ((\mathcal{D} * {W}^{(t)}) \otimes \mathcal{M}^{(t)} + {X}^{(t-1)} - {Y})$, and $\otimes^{T}$ denotes transposed convolution operator. Further, $\operatorname{prox}_{\theta_{ W}^{(t)}}$, $\operatorname{prox}_{\theta_{\mathcal{M}}^{(t)}}$ and $\operatorname{prox}_{\theta_{ X}^{(t)}}$ are three proximal operator networks.

\begin{table*}
\centering\setlength{\tabcolsep}{9pt}
\caption{Average PSNR/SSIM of different competing methods on synthesized DeepLesion~\cite{yan2018deep}.}\vspace{-2mm}
% $^{\text{*}}$ means we adopt the pre-trained model released by the authors \cite{wei2019semi}.
%\begin{tabular}{@{}c|c@{}c|c|c|c|c|c|c@{}}
% \tiny
% \setlength{\tabcolsep}{3.7pt}
\begin{tabular}{lcccccc}
    \toprule
Method & \multicolumn{5}{c}{ Large Metal \quad \quad   \quad\quad  $\longrightarrow$ \quad \quad Medium Metal \quad \quad $\longrightarrow$    \quad   \quad        Small Metal}                & Average      \\
\cmidrule(r){1-1} \cmidrule(r){2-6} \cmidrule(r){7-7}
Input & 24.12 / 0.6761 & 26.13 / 0.7471 & 27.75 / 0.7659 & 28.53 / 0.7964 & 28.78 / 0.8076 & 27.06 / 0.7586             \\
LI~\cite{kalender1987reduction} & 27.21 / 0.8920 & 28.31 / 0.9185 & 29.86 / 0.9464 & 30.40 / 0.9555 & 30.57 / 0.9608 & 29.27 / 0.9347   \\
NMAR~\cite{meyer2010normalized} & 27.66 / 0.9114 & 28.81 / 0.9373 & 29.69 / 0.9465 & 30.44 / 0.9591 & 30.79 / 0.9669 & 29.48 / 0.9442     \\
CNNMAR~\cite{zhang2018convolutional} & 28.92 / 0.9433 & 29.89 / 0.9588 & 30.84 / 0.9706 & 31.11 / 0.9743 & 31.14 / 0.9752 & 30.38 / 0.9644   \\
DuDoNet~\cite{lin2019dudonet} & 29.87 / 0.9723 & 30.60 / 0.9786 & 31.46 / 0.9839 & 31.85 / 0.9858 & 31.91 / 0.9862 & 31.14 / 0.9814   \\
DSCMAR~\cite{yu2020deep} & 34.04 / 0.9343 & 33.10 / 0.9362 & 33.37 / 0.9384 & 32.75 / 0.9393 & 32.77 / 0.9395 & 33.21 / 0.9375 \\
InDuDoNet~\cite{wang2021indudonet}& {36.74 / 0.9742} &{39.32 / 0.9893} & {41.86 / 0.9944} &{44.47 / 0.9948} & {45.01 / 0.9958} & {41.48 / 0.9897}\\
\cmidrule(r){1-1} \cmidrule(r){2-6} \cmidrule(r){7-7}
ACDNet \cite{wang2021dicdnet} & {37.28 / 0.9845} & {39.60 / 0.9903} & {42.19 / 0.9943} & {44.03 / 0.9957} & {44.11 / 0.9961} & 41.44 / 0.9922 \\
\tr{ACDNet-$p4$}&\tr{38.26 / 0.9900 }&\tr{40.36 / 0.9936}&\tr{42.35 / 0.9955}&\tr{44.93 / 0.9967}&\tr{45.21 / 0.9969}&\tr{42.22 / 0.9946} \\
\tr{ACDNet-$p8$} & \textbf{38.61} / \textbf{0.9893}  & {40.65 / 0.9930} & {42.73 / 0.9955 } & {\textbf{44.76} / \textbf{0.9967}} & {44.57 / 0.9968} & 42.26 / 0.9943 \\
\tr{ACDNet-$p8+$ }&\tr{38.17 / 0.9895 }&\tr{\textbf{41.73} / \textbf{0.9949}}&\tr{\textbf{43.13} / \textbf{0.9964}}&\tr{44.52 / 0.9972}&\tr{\textbf{44.61} / \textbf{0.9973}}&\tr{\textbf{42.43} / \textbf{0.9951}} \\
\cmidrule(r){1-1} \cmidrule(r){2-6} \cmidrule(r){7-7}
DICDNet \cite{wang2021dicdnet} & {38.08 / 0.9872}  & {39.98 / 0.9916}  & {42.64 / 0.9944} & {44.96 / 0.9955} & {45.55 / 0.9960} & {42.24 / 0.9930} \\
\tr{DICDNet-$p4$ }&\tr{38.80 / 0.9887 }&\tr{40.11 / 0.9922 }&\tr{43.00 / 0.9952 }&\tr{45.40 / 0.9961 }&\tr{46.03 / 0.9966}&\tr{42.67} / {0.9938}\\
\tr{DICDNet-$p8$ }& {38.40 / 0.9884}  & {40.16 / 0.9924} & {42.72 / 0.9950} & {45.39 / 0.9961} & {45.75 / 0.9965} & {42.48} / {0.9937} \\
\tr{DICDNet-$p8+$ }&\tr{\textbf{39.54} / \textbf{0.9907}}&\tr{\textbf{41.46} / \textbf{0.9939}}&\tr{\textbf{43.73} / \textbf{0.9960}}&\tr{\textbf{46.38} / \textbf{0.9970}}&\tr{\textbf{46.52} / \textbf{0.9971}}&\tr{\textbf{43.52} / \textbf{0.9949}} \\
\cmidrule(r){1-1} \cmidrule(r){2-6} \cmidrule(r){7-7}
OSCNet \cite{wang2022orientation} & {38.01 / 0.9871} & {40.20 / 0.9918} & {42.79 / 0.9945} & {44.51 / 0.9951} & {\textbf{45.55} / \textbf{0.9960}} & 42.21 / 0.9929 \\
\tr{OSCNet-$p4$ }&\tr{38.59 / 0.9889}&\tr{40.07 / 0.9923}&\tr{42.92 / 0.9952}&\tr{\textbf{45.55} / \textbf{0.9962}}&\tr{46.02 / 0.9967}&\tr{42.63} / {0.9938}\\
\tr{OSCNet-$p8$ }& {38.60 / 0.9891} & {40.18 / 0.9926} & {42.71 / 0.9951} & {44.99 / 0.9959} & {45.54 / 0.9965} & {42.40} / {0.9938} \\
\tr{OSCNet-$p8+$ }&\tr{\textbf{39.25} / \textbf{0.9897}}&\tr{\textbf{40.44} / \textbf{0.9928}}&\tr{\textbf{43.25} / \textbf{0.9955}}&\tr{45.29 / 0.9961}&\tr{\textbf{46.26} / \textbf{0.9969}}&\tr{\textbf{42.90} / \textbf{0.9942}}\\
    \bottomrule
\end{tabular}
\label{tab:tabmar}
\vspace{-2mm}
\end{table*}

\begin{figure*}[t]
    \centering\hspace{5mm}
    \includegraphics[width=0.85\linewidth]{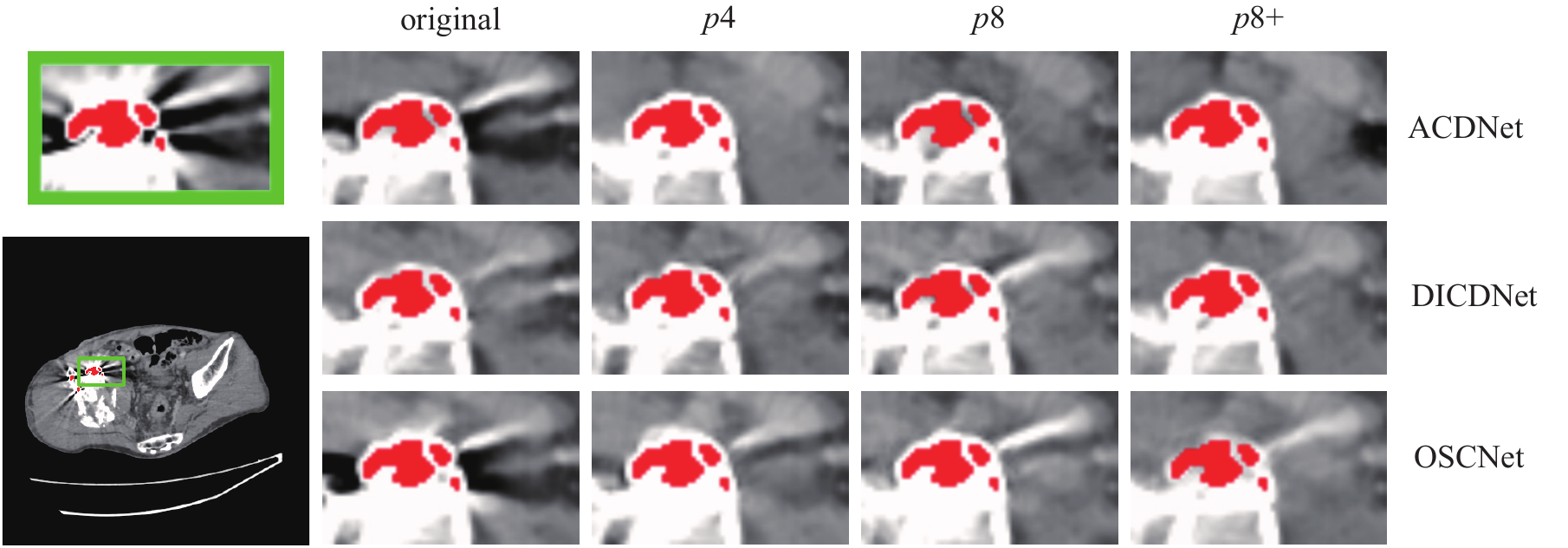}\vspace{-0mm}
    % \subfigure[Jackson Yee]{\includegraphics[width=3.5cm]{image/014GT.png}}
    % \subfigure[Jackson Yee]{\includegraphics[width=3.5cm]{image/014GT.png}}
    \caption{Generalization results on a real clinical metal-affected CT image from CLINIC-metal \cite{liu2021deep}. The red pixels stand for metallic implants, which are segmented with the thresholding of 2500 HU.} %图片标题
    \label{fig:mar_clinic}  %图片交叉引用时的标签
 \vspace{-2mm}
\end{figure*}

\vspace{1mm}
It is evident that both the human tissue structures and the metal artifacts inherently exhibit rotation symmetry prior, as visually demonstrated in Fig. \ref{fig:mar01}. Therefore, it is rational to employ rotation equivariant proximal networks for the estimation of  $X$ within the frameworks of ACDNet, DICDNet and OSCNet.  

% \subsection{Network architecture settings}
% \vspace{2mm}
{\bf Network architecture setting.}
\tr{Similar to the experiments in Section 4.1,  we construct equivariant proximal network by replacing the regular convolution in the original proximal network with equivariant convolution, F-Conv \cite{xie2022fourier}, and test the performance under 3 typical settings. Specifically, for ACDNet, we construct ACDNet-$p4$, ACDNet-$p8$ and ACDNet-$p8$+. Among these 3 methods,  the  tag ``-$p4$'' and ``-$p8$'' indicate $t = 4$ and $t = 8$, respectively.  Besides, the channel numbers of these two methods are set as $\sfrac{1}{4}$ and $\sfrac{1}{8}$ of the original non-equivariant network respectively. The tag ``-$p8$+''  indicates that $t = 8$ and  $\sfrac{1}{4}$ channel number of the original network, whose parameter number in each convolution is the same with ``-$p4$'' case.   Moreover, we use the same notation strategy for the enhanced DICDNet and OSCNet, ensuring clarity and coherence throughout.
The  $3 \times 3$ filter in the original network is replaced with the $5 \times 5$ filter in F-Conv for better rotating the filter. All the training settings and loss function are set the same as the original methods for fair competition.}

% \subsection{Datasets and Training Settings}
{\bf Datasets and Training Settings.}
Following the synthetic procedure in \cite{wang2021dicdnet}, \cite{wang2022orientation}, we can generate the paired $ X$ and $ Y$ for training and testing by using 1,200 clean CT images from DeepLesion \cite{yan2018deep} and 100 simulated metallic implants from \cite{zhang2018convolutional}. Consistent with the original work, we randomly select 90 masks and 1000 clean CT images to synthesize metal-corrupted training samples. The remaining 10 masks and 200 clean CT images are used to generate test samples. For training, except that the total epoch of DICDNet is changed to 200 to ensure convergence, all methods of replacing the rotation equivariant proximal network remain consistent with the original method.

% \subsection{Quantitative and Qualitative Comparison}
{\bf Quantitative and Qualitative Comparison.}
\tr{As illustrated in Table \ref{tab:tabmar}, it is evident that when replacing the proximal networks with its rotation equivariant version,  the performance is significantly enhanced and outperforms the competing methods. Specifically,  the method with tag ``-$p8$+'' seems to achieve the best overall performance; the methods  tag ``-$p8$'' and tag ``-$p4$'' demonstrate comparable performance, while ``-$p8$'' method contains much fewer parameters.} Visually, as demonstrated in Fig. \ref{fig:mar_deeplesion}, the method with the replaced rotation equivariant proximal network performs better than the original method in removing metal artifacts and restoring the contour of human tissue structures. Both quantitative and qualitative comparisons reveal the superiority of the rotation equivariant embedded network over the original one. %\tr{Visualization of feature maps is available in the supplementary material.}

%----------------------------
\begin{table*}
% \vspace{-1mm}
\centering
\caption{Average PSNR/SSIM of different competing methods on four benchmark datasets.}

\setlength{\tabcolsep}{16pt}
\begin{tabular}{ccccccccc}
    \toprule
  \multirow{2}{*}{Method} & \multicolumn{2}{c}{Rain100L \cite{yang2019joint}} & \multicolumn{2}{c@{}}{Rain100H \cite{yang2019joint}} & \multicolumn{2}{c}{Rain1400 \cite{fu2017removing}} & \multicolumn{2}{c@{}}{Rain12 \cite{li2016rain}}\\%& Fig.\ref{e1} & Fig.\ref{e2}& Fig.\ref{e1} &
\cmidrule(r){2-3}\cmidrule(r){4-5}\cmidrule(r){6-7}\cmidrule(r){8-9}
   & PSNR & SSIM & PSNR & SSIM  & PSNR & SSIM & PSNR & SSIM\\
    \midrule
  Input & 26.90 & 0.8384 & 13.56 & 0.3709 & 25.24 & 0.8097 & 30.14 & 0.8555\\

  DSC\cite{luo2015removing} & 27.34 & 0.8494 & 13.77 & 0.3199  & 27.88 &0.8394 & 30.07 &0.8664\\

  GMM\cite{li2016rain} &29.05 &0.8717 & 15.23 &0.4498  &27.78 & 0.8585 & 32.14 & 0.9145 \\

  JCAS\cite{gu2017joint}  & 28.54 & 0.8524 & 14.62 & 0.4510 &26.20 & 0.8471 & 33.10 &0.9305 \\

  Clear\cite{fu2017clearing} &30.24 & 0.9344 & 15.33 & 0.7421 & 26.21& 0.8951 & 31.24 & 0.9353\\

  DDN\cite{fu2017removing}& 32.38 & 0.9258 & 22.85 & 0.7250 & 28.45 & 0.8888 & 34.04 & 0.9330 \\
%\Xhline{0.5pt}
%  JORDER\cite{Yang2019Joint}$^{\text{o}}$ (CVPR'17) & 36.61 & 0.9740 & 26.54 & 0.8350 \\

  RESCAN\cite{li2018recurrent}   & 38.52& 0.9812 &29.62 & 0.8720 &32.03& 0.9314 &36.43&0.9519\\

  PReNet\cite{ren2019progressive}& 37.45& 0.9790 &30.11& 0.9053 & 32.55 & 0.9459 & 36.66& 0.9610\\

  SPANet\cite{wang2019spatial} & 35.33 & 0.9694 &25.11 & 0.8332 & 29.85& 0.9148 & 35.85& 0.9572 \\

  JORDER\_E\cite{yang2019joint} & 38.59 & 0.9834 & 30.50 &0.8967 &32.00 & 0.9347 & 36.69 & 0.9621\\

  SIRR\cite{wei2019semi} & 32.37 & 0.9258 & 22.47 & 0.7164 & 28.44 & 0.8893 & 34.02& 0.9347\\
  \tr{IDT\cite{xiao2022image}} & \tr{35.42} & \tr{0.9674} & \tr{30.45} & \tr{0.9081} & \tr{\bf 33.55} & \tr{\bf 0.9531} & \tr{35.98} & \tr{0.9584} \\
    \midrule
RCDNet\cite{wang2020model} & 40.00 & 0.9860 & 31.28 & 0.9093 & 33.04 & 0.9472 & 37.61 & 0.9644 \\

  \tr{RCDNet-$p4$} & \tr{40.19} & \tr{0.9862} & \tr{31.69} & \tr{0.9134} & \tr{33.23} & \tr{0.9487} & \tr{\bf37.97} & \tr{ \bf 0.9651} \\
  \tr{RCDNet-$p8$} & 40.28 & 0.9867 & 31.43 & 0.9108 & 33.22 & 0.9488 & 37.66 & 0.9617 \\
  \tr{RCDNet-$p8+$} & \tr{\bf 40.52} & \tr{\bf 0.9873} & \tr{\bf 31.81} & \tr{\bf 0.9151} & \tr{33.39} & \tr{0.9503} & \tr{37.73} & \tr{0.9626} \\
    \bottomrule
\end{tabular}\vspace{-2mm}
\label{tab:tabrain}
\end{table*}
%------------------------------

 {\bf Generalization Results.}
In order to show the generalization advantage of the rotation equivariant embedded method, we utilize another public dataset, CLINIC-metal \cite{liu2021deep}, for testing. As shown in Fig.~\ref{fig:mar_clinic}, the rotation equivariant proximal methods outperform the original results in terms of shadings and streaking artifacts removal, and they more accurately recover the human tissue structures. This provides an intuitive demonstration of the improved generalization ability of the ameliorated method.

\subsection{Single Image Rain Removal}\label{RCD_exp}
Images captured in rainy scenes suffer from unfavorable visual degradation, which increases the difficulty of many outdoor computer vision tasks, such as automatic driving and video surveillance. Therefore, rain removal from images is a meaningful research problem. Recently, Wang et al.
designed a deep unfolding model, RCDNet \cite{wang2020model}, for rain removal based on a convolutional dictionary model and achieved SOTA performance. Normally, the observed color rainy image is denoted as $\mathcal{O} \in \mathbb{R}^{H \times W \times 3}$, which can be regarded as composed of the clean background image and the rain layer, expressed as follows:
\begin{equation}
    \mathcal{O} = \mathcal{B} + \mathcal{R}=\mathcal{B} + \sum\limits_{n=1}^{N} \mathcal{C}_n \otimes M_n,
    \label{eq:eq19}
\end{equation}
where $\mathcal{B}$ and $\mathcal{R}$ denote the background and the rain layers, respectively, $\mathcal{C}_n$ represents convolutional dictionary terms and $M_n$ represents the corresponding rain feature map.

%-----------------------------------------------
\begin{figure*}[ht]
    \centering
    \includegraphics[width=18cm]{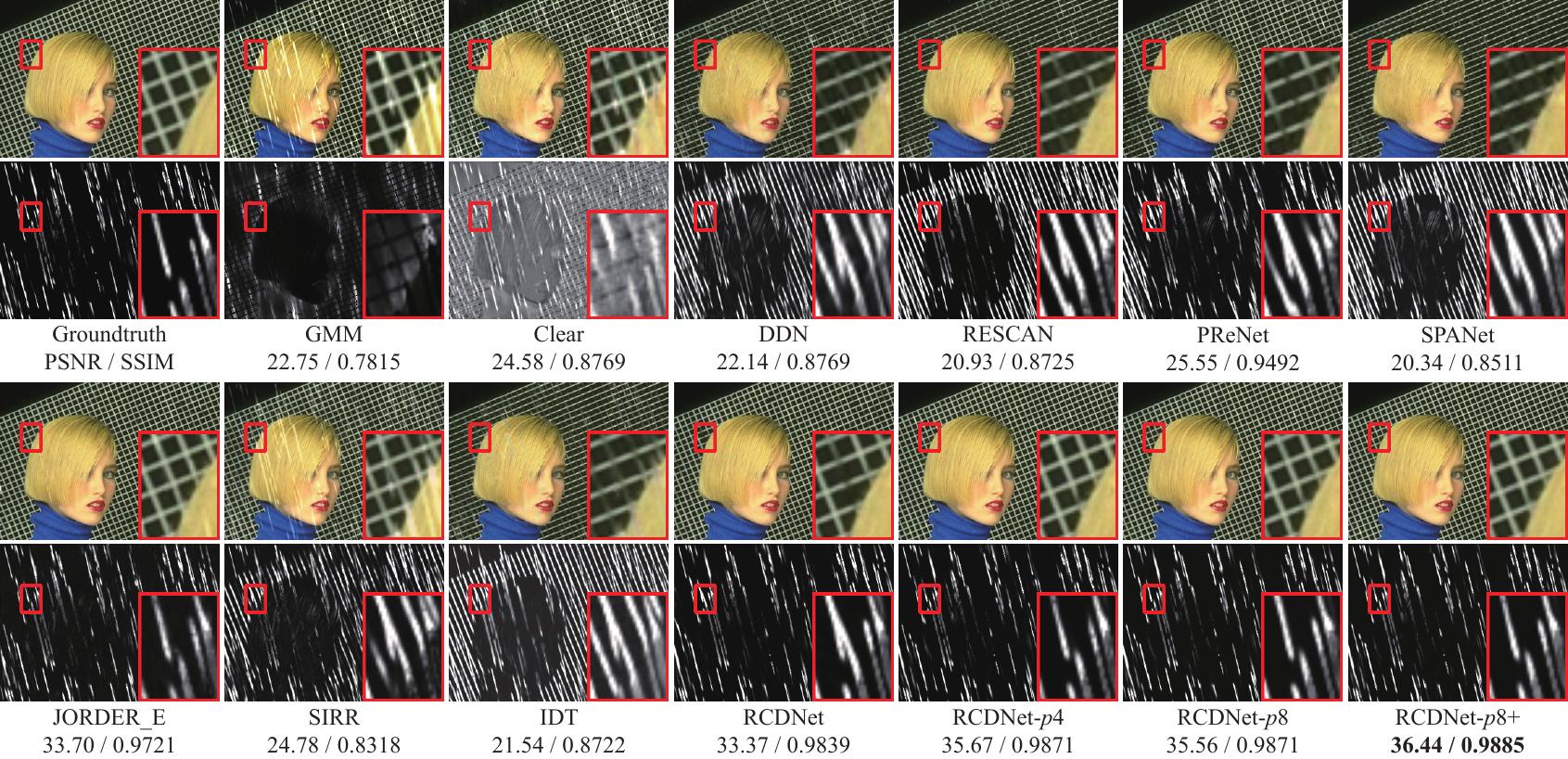}
    \caption{The $1^{st}$ column: a typical ground truth sample in Rain100L \cite{yang2019joint} dataset (upper) and its ground truth rain layer (lower). The $2^{nd}-12^{th}$ columns: derained results (upper) and extracted rain layers (lower) by all competing methods}
    \label{fig:derain}
    \vspace{-2mm}
\end{figure*}
%------------------------------------------------

%-------------------------------------
\begin{table*}[t]
    \centering
    \caption{Average PSNR and SSIM comparisons on Dense10\cite{wei2019semi} and Spare10\cite{wei2019semi} of different deep models trained on Rain100H.}
    \setlength{\tabcolsep}{5.5pt}
    \begin{tabular}{c c c c c c c c c c c c c}
    \toprule
    Datasets & Metrics & Input & DSC & JCAS  & DDN & RESCAN & PReNet & SPANet & JORDER\_E & SIRR & RCDNet & RCDNet-$p8$ \\
    \midrule
    \multirow{2}{*}{Dense10} & PSNR & 19.17 & 20.85 & 19.93 & 21.38 & 21.81 & 21.91 & 20.28 & 21.66 & 21.23 & 21.89 & {\bf 22.03} \\
      & SSIM & 0.8495 & 0.8811 & 0.8694 & 0.8965 & 0.9073 & 0.9236 & 0.8982 & 0.9093 & 0.8925 & 0.9239 & {\bf 0.9248} \\
    \midrule
    \multirow{2}{*}{Spare10} & PSNR & 25.42 & 26.37 & 26.38 & 27.83 & 28.73 & 29.02 & 27.97 & 27.66 & 27.48 & 29.69 & {\bf 29.75} \\
      & SSIM & 0.8956 & 0.8989 & 0.9043 & 0.9249 & 0.9337 & 0.9412 & 0.9279 & 0.9257 & 0.9181 & 0.9422 & {\bf 0.9422} \\
    \bottomrule
    \end{tabular}
    \label{tab:derain_generlization}\vspace{-2mm}
\end{table*}
%-------------------------------------

RCDNet exploits the proximal gradient descent algorithm and its unfolding network to estimate rain maps $\mathcal{M}_n$ and rain-free background $\mathcal{B}$, simultaneously. It can also be classified as instances of the broader inverse problem (\ref{eq:eq1}). For solving $\mathcal{M}$ and $\mathcal{B}$, by manners similar to Eqs. (\ref{eq:ista})-(\ref{eq:ista_prox}), the constructed network structures as follows:
%For convenience, we write the summation as a tensor form $\mathcal{C} \otimes M$, where $\mathcal{C} \in \mathbb{R}^{k \times k \times 3 \times N}$ and $\math$
\begin{equation}
    \left \{
    \begin{aligned}
        \mathcal{M}^{(t)} & = \operatorname{prox}_{\theta_{\mathcal{M}}^{(t)}} (\mathcal{M}^{(t-1)} - \eta_1 \nabla g_1(\mathcal{M}^{(t-1)})) \\
        \mathcal{B}^{(t)} & = \operatorname{prox}_{\theta_{\mathcal{B}}^{(t)}} (\mathcal{B}^{(t-1)} - \eta_2 \nabla g_2(\mathcal{B}^{(t-1)}))
    \end{aligned}
    \right.
    \label{eq:eq22}
\end{equation}
where $\nabla g_1(\mathcal{M}^{(t-1)})) = \mathcal{C} \otimes^{T} ( \sum_{n=1}^{N} \mathcal{C}_n \otimes M_n^{(t-1)} + \mathcal{B}^{(t-1)} - \mathcal{O} )$, $\nabla g_2(\mathcal{B}^{(t-1)}) = \sum_{n=1}^N \mathcal{C}_n \otimes M_n^{(t)} + \mathcal{B}^{(t-1)} - \mathcal{O}$, $\mathcal{C} \in \mathbb{R}^{k\times k \times N \times 3}$ is a 4-dimensional tensor stacked by $\mathcal{C}_n, n = 1,2,\cdots,N$; $\otimes^{T}$ denotes the transposed convolution, $\operatorname{prox}_{\theta_{\mathcal{M}}^{(t)}}(\cdot)$ and $\operatorname{prox}_{\theta_{\mathcal{B}}^{(t)}}(\cdot)$ are two proximal networks consisting of a series of residual blocks, respectively. All network parameters can be learned in an end-to-end manner.

Since rain-free images have rich rotation symmetry structures and textures, the restored background image should maintain this structure prior even after rain streak removal. Therefore, it is rational to construct a rotation equivariant proximal network for $\mathcal{B}$. 

{\bf Network architecture setting.}
\tr{The equivariant proximal network involves four CNN based residual blocks.
Similar as the experiments in Section 4.1,  we construct RCDNet-$p4$, RCDNet-$p8$ and RCDNet-$p8$+ to verify the effectiveness of rotation equivariant proximal networks, with setting $t = 4$ \& $\sfrac{1}{4}$ channel number, $t = 8$ \& $\sfrac{1}{8}$ channel number and $t = 8$ \& $\sfrac{1}{4}$ channel number, respectively.
The  $3 \times 3$ filter in the original network is replaced with the $5 \times 5$ filter in F-Conv for better rotating the filter. And all the training settings and loss function are set the same as the original methods for fair competition.}

 We exploit the $p8$ group for the equivariant convolutions and set the channel number of each residual block as $\sfrac{1}{8}$ to the original networks in each layer to keep their similar memory. The $3 \times 3$ size filter in the original network is replaced with the $5 \times 5$ filter in F-Conv to preserve the number of parameters. All the training settings and loss function are set the same as the original RCDNet for fair competition.

{\bf Datasets and Training settings.}
We compare our method with typical single image rain removal SOTA methods, including DSC~\cite{luo2015removing}, GMM~\cite{li2016rain}, JCAS~\cite{gu2017joint}, Clear~\cite{fu2017clearing}, DDN~\cite{fu2017removing}, RESCAN~\cite{li2018recurrent}, PReNet~\cite{ren2019progressive}, SPANet~\cite{wang2019spatial}, JORDER\_E~\cite{yang2019joint}, SIRR~\cite{wei2019semi}, and RCDNet on four commonly-used benchmark datasets, i.e., Rain100L \cite{yang2019joint}, Rain100H \cite{yang2019joint}, Rain1400 \cite{fu2017removing}, and Rain12~\cite{li2016rain}. The training strategy is carried out according to the original settings, and the strategy of RCDNet-$\text{E}_{\mathcal{B}}$ is the same as RCDNet.

% \subsection{Quantitative and Qualitative Comparison}
{\bf Quantitative and Qualitative Comparison.} \tr{As shown in Table \ref{tab:tabrain},  with the proposed rotation equivariant proximal,  RCDNet-$p4$, RCDNet-$p8$ and RCDNet-$p8$+ consistently outperform RCDNet, and achieve comparable performance with the transformer-based SOTA method IDT.  Particularly, in the datasets Rain100L, Rain100H and Rain12, the RCDNet-$p8$+ method achieves the best performance among these methods. Besides, it should be noted that the parameter of RCDNet-$p8$ is much less than IDT\footnote{Please refer to the supplementary material for a comparison between RCDNet-$p8$+ and IDT, when increasing the parameter number of RCDNet-$p8$+ to the same level as IDT.}.  Furthermore, as depicted in Fig. \ref{fig:derain}, our proposed method exhibits superior performance in removing visual rain streaks. It is worth noting that the model based on convolution sparse coding often mistakenly identifies white stripes as rain stripes, while RCDNet configured with rotation equivariant proximal network can effectively alleviate this shortcoming.
These results further support the effectiveness of adopting the rotation equivariant proximal network in this task.} 
%\tr{Visualization of feature maps is available in the supplementary material.}

{\bf Generalization Results.}
We then evaluate the proposed method in the case that rain types are inconsistent between training and testing. We adopt Dense10 \cite{wei2019semi} and Sparse10 \cite{wei2019semi} to evaluate the generalization capability of all competing methods trained on Rain100H. From the quantitative results depicted in Table \ref{tab:derain_generlization}, it is seen that RCDNet-$\text{E}_{\mathcal{B}}$ obtains higher PSNR and SSIM. This verifies that the proper embedding of rotation equivariance prior knowledge into the network enhances its generalization ability.

\vspace{-2mm}
\subsection{Comparison of Number of Parameters}\label{num_parameter}
When using filter with the same size, the parameter number in each rotation equivariant convolution is  $\sfrac{1}{t}$ of that in regular convolution. However, rotation equivariant convolution methods usually exploit larger filter for better rotating the filter. It's thus necessary to count and compare the parameter number of the original networks and the enhanced networks. Specifically, in 2 scale image super-resolution, the parameter is 6.50M, 5.50M, 4.12M and 8.47M for original KXNet, KXNet-$p4$, KXNet-$p8$ and KXNet-$p8$+, respectively. Since KXNet-$p8$  performs better than KXNet, we can conclude that rotation equivariant proximal network can achieve better results even it is more lightweight. Similar phenomenon can also be observed in the other two experiments. Specifically,  the parameter is 1.60M and 1.27M for  RCDNet and RCDNet-$p8$, respectively, it's 3.17M and 2.40M for  RCDNet and RCDNet-$p8$, respectively.
For more details, please refer to the supplementary material.

\vspace{-3mm}
\section{Conclusion}
In this study, we have suggested a rotation equivariant proximal network based on F-Conv to alleviate the limitations of existing standard CNN-based proximal networks in capturing the rotation symmetry prior to images. Our approach embeds rotation symmetry priors into the deep unfolding framework via a rotation equivariant proximal network, leading to a more accurate representation of the image prior and improved model generalization ability. By leveraging the fact that the proximal network is rotation equivariant when the regularization term is invariant to rotation transformations, we have demonstrated the effectiveness of our proposed method in multiple typical low-level vision tasks, including blind image super-resolution, medical CT image reconstruction, and image de-raining, outperforming standard CNN-based proximal networks in capturing the rotation symmetry prior of images. Especially, we have first provided the theoretical equivariant error for such a designed proximal network with multi-layer networks under arbitrary rotation degrees, which should be the most refined theoretical conclusion for such error evaluation nowadays. This is specifically indispensable for supporting the rationale of this line of deep unfolding networks with intrinsic interpretability requirements and the potential usefulness of this methodology on more extensive image restoration tasks.

There is still large room for further enhancing the performance of the equivariant proximal operators. Firstly, proximal operators with more transformation equivariance are worth exploring, such as scale, affine, and color equivariance. Besides, equivariant modification for transformer-based proximal operators will also be explored in our future research. A more comprehensive and deep exploration of filter parametrization methods for designing high-accuracy equivariant convolution beyond F-Conv is also an important issue. Intuitively, bicubic/bilinear interpolation could be more suitable for filter parametrization in IR tasks, compared to the Fourier series expansion exploited in F-Conv.

%\section*{Acknowledgement}
%We want to sincerely thank the anonymous reviewers for their constructive comments and suggestions, which have helped significantly improve the quality of this paper.
%
%This research was supported by NSFC project under contracts U21A6005, 61721002, U1811461, 62076196, the Major Key Project of PCL under contract PCL2021A12, and the Macao Science and Technology Development Fund under Grant 061/2020/A2.

\bibliographystyle{unsrt}
%%\begin{thebibliography}{10}\itemsep=-1pt
%\bibliography{egbib}

% \vspace{-10mm}
\begin{IEEEbiography}[{\includegraphics[width=1in,height=1.25in,clip,keepaspectratio]{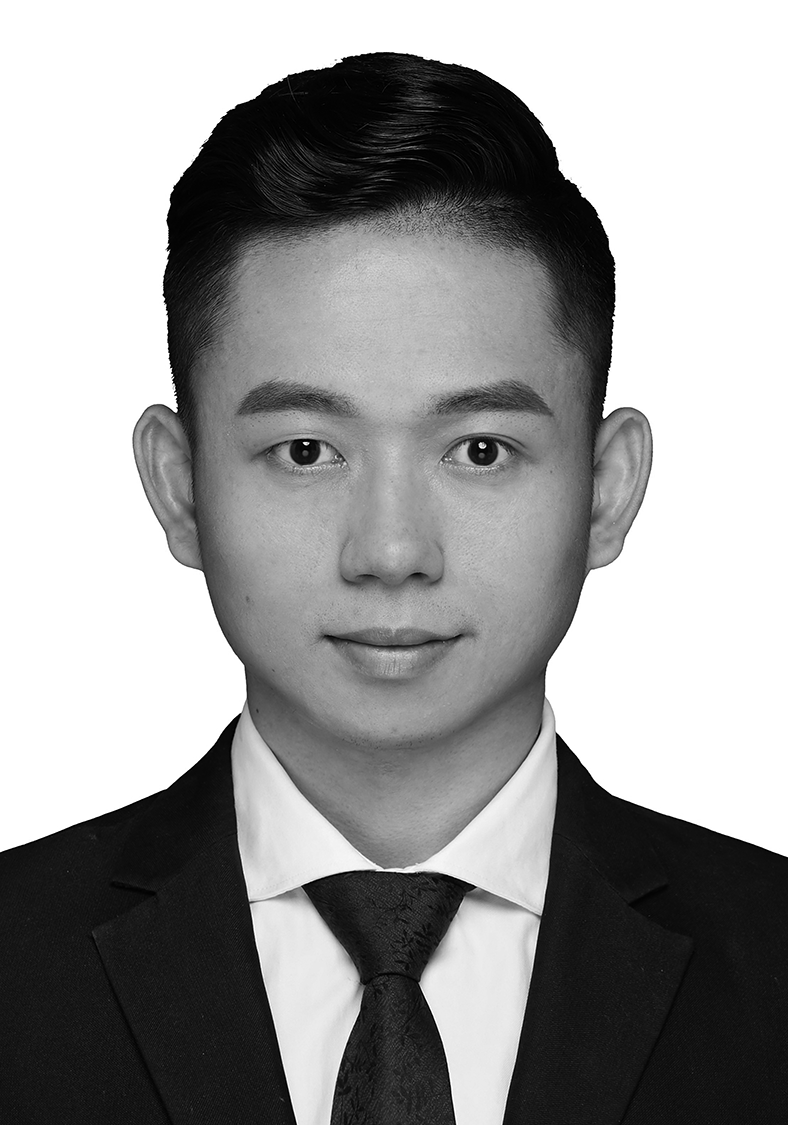}}]{Jiahong Fu} received the B.Sc. degree from North China Electric Power University, Beijing, China, in 2019. He is currently pursuing the Ph.D. degree in Xi'an Jiaotong University.
His current research interests include model-based deep learning and rotation equivariant deep learning.
%\vspace{-0mm}
\end{IEEEbiography}
\vspace{-7mm}
\begin{IEEEbiography}[{\includegraphics[width=1in,height=1.25in,clip,keepaspectratio]{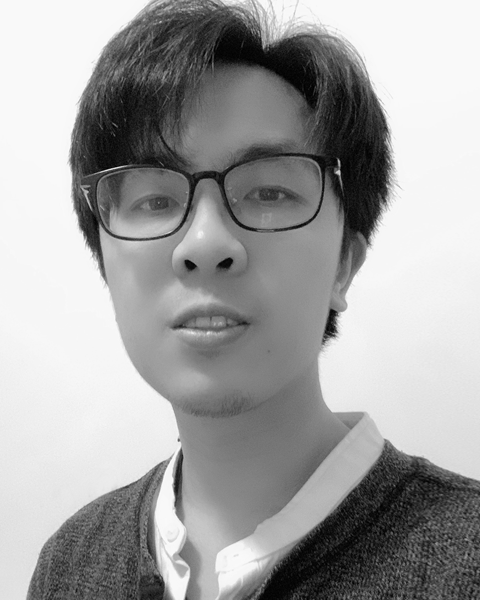}}]{Qi Xie} received the B.Sc. and Ph.D degree from Xi'an Jiaotong University, Xi'an, China, in 2013 and 2020 respectively.  Form 2018 to 2019, he was a Visiting Scholar in Princeton University, Princeton, NJ, USA. 
He is currently an associate professor in School of Mathematics and Statistics, Xi'an Jiaotong University.
His current research interests include model-based deep learning, filter parametrization-based deep learning, rotation equivariant deep learning.
%\vspace{-0mm}
\end{IEEEbiography}
\vspace{-7mm}
\begin{IEEEbiography}[{\includegraphics[width=1in,height=1.25in,clip,keepaspectratio]{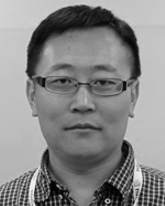}}]{Deyu Meng} received the B.Sc., M.Sc., and Ph.D. degrees from Xi'an Jiaotong University, Xi'an, China, in 2001, 2004, and 2008, respectively. He is currently a professor in School of Mathematics and Statistics, Xi'an Jiaotong University, and adjunct professor in Faculty of Information Technology, The Macau University of Science and Technology. From 2012 to 2014, he took his two-year sabbatical leave in Carnegie Mellon University. His current research interests include model-based deep learning, variational networks, and meta-learning.
%\vspace{-0mm}
\end{IEEEbiography}
\vspace{-7mm}
\begin{IEEEbiography}[{\includegraphics[width=1in,height=1.25in,clip,keepaspectratio]{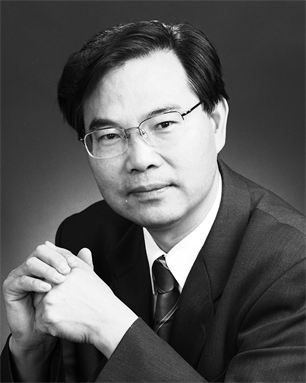}}]{Zongben Xu} received the Ph.D. degree in mathematics from Xi'an Jiaotong University, Xi'an, China, in 1987. He currently serves as the Academician of the Chinese Academy of Sciences, the Chief Scientist of the National Basic Research Program of China (973 Project), and the Director of the Institute for Information and System Sciences with Xi'an Jiaotong University. His current research interests include nonlinear functional analysis and intelligent information processing.
He was a recipient of the National Natural Science Award of China in 2007 and the winner of the CSIAM Su Buchin Applied Mathematics Prize in 2008.
\end{IEEEbiography}
\end{document}